\documentclass[aps,prb,longbibliography,groupedaddress,twocolumn,nofootinbib,10pt]{revtex4-2}
\usepackage{graphicx}
\usepackage{mathtools}
\usepackage{amsmath}
\usepackage{amssymb}
\usepackage{bbm}
\usepackage{bm}
\usepackage{cancel}
\usepackage{xspace}
\usepackage{braket}
\usepackage{xcolor}
\usepackage{siunitx}
\usepackage{calrsfs}
\usepackage{comment}
\usepackage{placeins}
\usepackage[export]{adjustbox}
\usepackage{wrapfig}
\usepackage{bbold}
\usepackage{booktabs} 
\usepackage[colorlinks=true]{hyperref} 

\newcommand{\ee}{\mathrm{e}}  
\DeclareMathOperator*{\ii}{i} 
\newcommand*\dd{\mathop{}\!\mathrm{d}}

\renewcommand{\vec}[1]{\bm{#1}} 
\newcommand{\kel}[1]{\underline{#1}} 

\definecolor{hblue}{RGB}{0,80,255}
\definecolor{hred}{RGB}{255,80,0}

\begin{document}

\title{Mixed-configuration approximation for multiorbital systems out of equilibrium}

\author{Tommaso Maria Mazzocchi}
\email[]{mazzocchi@tugraz.at}
\author{Daniel Werner}
\author{Markus Aichhorn}
\author{Enrico Arrigoni}
\email[]{arrigoni@tugraz.at}
\affiliation{Institute of Theoretical and Computational Physics, Graz University of Technology, 8010 Graz, Austria}

\date{\today}


\begin{abstract}

We propose a mixed-configuration approximation based on single-band impurity solvers to efficiently study nonequilibrium multi-orbital systems at moderate computational cost. In this work, we merge the approach with the so-called auxiliary master equation approach. As benchmark, we first show that our approach reproduces the results of quantum Monte Carlo (QMC) for two-orbital impurity models at equilibrium with overall good accuracy, especially for non-degenerate orbitals. We then use our approach as impurity solver for  dynamical mean-field theory (DMFT) to address the case of a two-orbital, realistic layered structure, recovering the strong crystal-field-driven charge polarization observed by solving the DMFT self-consistent cycle with QMC, albeit slightly reduced. Finally, we address a prototype nonequilibrium setup by sandwiching this layer between metallic contacts subject to a bias voltage described by different chemical potentials. This simplified model demonstrates our method’s potential to access nonequilibrium steady-state behavior of multi-orbital, realistic materials. These findings provide a first-step basis for theoretical studies of nonequilibrium properties of multi-orbital compounds directly in the real frequency domain.
\end{abstract}

	
\maketitle

\section{Introduction}\label{sec:intro}

Predicting the nonequilibrium properties of real materials has garnered increasing attention over the past few decades. Strongly correlated materials, in particular, remain a central focus due to their rich phase diagram. For example, the so-called equilibrium insulator-to-metal transition can be driven by several factors---including pressure-induced strain, temperature, and doping, see Ref.~\cite{ja.tr.15} for an extensive, though not exhaustive, review---which are often associated with a resistive switch. 
From a modelling prespective, dynamical mean-field theory (DMFT)~\cite{me.vo.89,ge.ko.92,ge.ko.96} has become a powerful method for describing both strongly correlated lattice systems and realistic materials, especially when employed in combination with density functional theory (DFT). One prominent example of the application of DFT+DMFT in equilibrium is SrVO$_3$, see, e.g., Refs.~\cite{lieb.03,se.fu.04,pa.bi.04.po,ne.ke.05,ne.he.06,no.ka.12}. This material is widely regarded as the paradigmatic {\em correlated metal} because in its bulk form it remains metallic with a moderately enhanced effective mass $m^*/m\approx 2$, even when chemically substituted, as is the case of Ca$_{1-x}$Sr$_x$VO$_3$~\cite{in.go.98}. Intriguingly, when SrVO$_3$ is grown as a very thin film of a few monolayers thickness, the compound becomes insulating---a phase that can persist even in structures as thin as three monolayers~\cite{yo.ok.10,zh.wa.15}. Unlike many layered materials where charge transfer primarily occurs at interfaces between different layers, the crystal field in layered SrVO$_3$ confines charge redistribution to the orbitals within the same monolayer. Consequently, a pronounced charge polarization develops: the \(d_{xy}\) orbital of the \(t_{2g}\) manifold becomes almost fully occupied, while the other two orbitals (\(d_{yz}, d_{zx}\)) remain nearly empty~\cite{ba.as.16}.

The theoretical investigation of nonequilibrium setups for this compound could yield valuable insights to design possible experiments. As a first step one might examine, for instance, the conducting properties of layered SrVO$_3$ by sandwiching it between two metallic contacts and applying a bias voltage. To this end, it is necessary to employ the nonequilibrium extension of DMFT, which has been successfully applied to various prototype systems out of equilibrium, including Mott–Hubbard models subject to external driving in the fashion of static and/or periodic electric field~\cite{mu.we.18,pi.li.21,ma.ga.22,ga.ma.22,ha.ar.23,ga.we.24} as well as disordered infinite-dimensional lattices with or without interaction quenches~\cite{do.te.22,ma.we.23}, only to mention a few examples from recent years.

The DMFT approximation to lattice models in and out of equilibrium constitutes a cornerstone of theoretical solid-state physics~\cite{ge.ko.96,ao.ts.14}. It maps the lattice model onto a single-impurity model coupled to a bath that is to be determined self-consistently. Even in equilibrium, solving the so-called impurity problem represents a major computational bottleneck, particularly when using numerically exact methods such as quantum Monte Carlo (QMC)~\cite{gu.ja.91}, the numerical renormalization group (NRG)~\cite{wils.75}, or exact diagonalization techniques like the so-called auxiliary master equation approach (AMEA)~\cite{do.nu.14,we.lo.23}. In addition, under nonequilibrium conditions, few impurity solvers retain their accuracy without a prohibitive computational cost so that they can be efficiently used for DMFT, where the solution of the impurity problem has to be solved multiple times.

Investigating realistic materials within the DMFT framework requires multi-orbital impurity solvers, which comes at a certain cost. Depending on the details of the algorithm, this can be severe fermionic sign problems as for interaction bases series expansions in QMC~\cite{gu.mi.11}. Another quite common problem is the exponential growth of Hilbert space with system size, which makes these numerical methods exponentially costly, see for instance Refs.~\cite{li.is.11,cr.kr.25,mi.ga.14,st.mi.16,we.co.06,ai.bi.10,gu.mi.11,yi.ha.11,ei.gu.20}. Although there have been recent advances, for instance using tensor-train techniques~\cite{nu.je.22}, these issues are not been resolved yet.
In nonequilibrium, QMC can address single-impurity problems for short to medium time evolutions~\cite{co.gu.15}, and even address steady-state directly~\cite{er.gu.23} within the inchworm algorithm, which recently has been applied to multiorbital systems~\cite{er.bl.24}. 

In this work, we introduce an approximate scheme to tackle the multi-orbital problem, in particular in nonequilibrium steady-state settings. The method maps the multi-orbital impurity problem onto a set of independent, different single-impurity problems. Each impurity problem is solved separately for each orbital under fixed configurations of the others, and the solutions of the individual impurities are combined using 
appropriate configuration probabilities, which are determined self-consistently. 
This \emph{mixed-configuration approximation} (MCA) can tackle multi-orbital systems in and out of equilibrium at moderate computational cost. Our approach treats intra-orbital interactions exactly while approximating inter-orbital effects in a mean-field–like fashion. This scheme can be based on any single-band impurity solver that would be hard to extend beyond one orbital.
In this work, we employ AMEA as a single-impurity solver, which is particularly suited as a many-body impurity solver for nonequilibrium steady-state problems.

We first benchmark our MCA-AMEA scheme against QMC for various two-orbital impurity models in equilibrium. Then we apply it to realistic lattice systems, investigating a toy-model lattice system inspired by the SrVO$_3$ layer of Ref.~\cite{ba.as.16}, both in and out of equilibrium, within the DMFT framework.

The remainder of this work is organized as follows. In Sec.~\ref{sec:model}, we introduce the model Hamiltonian while in Sec.~\ref{sec:method} we detail our mixed-configuration approximation using a two-orbital impurity model as an illustrative example. Section~\ref{sec:results} discusses our results for both the impurity models and the correlated lattice together with the application to a feasible nonequilibrium setup. Finally, Sec.~\ref{sec:conclusions} offers concluding remarks and outlines further applications, including realistic systems with more than two orbitals both in and out of equilibrium.

\section{Multi-orbital impurity model}\label{sec:model}

The Hamiltonian for a multi-orbital impurity model coupled to a fermionic bath is given by
\begin{equation}\label{eq:H_tot}
\hat{H} = \hat{H}_{\text{imp}} + \hat{H}_{\text{bath}} + \hat{H}_{\text{i-b}},
\end{equation}
where $\hat{H}_{\text{imp}}$ can be split as
\begin{equation}\label{eq:H_imp}
\hat{H}_{\text{imp}} = \hat{H}_{0} + \hat{H}_{\text{int}}.
\end{equation}
The noninteracting Hamiltonian $\hat{H}_{0}$ in Eq.~\eqref{eq:H_imp} reads
\begin{equation}\label{eq:nonint_Hamiltonian}
\hat{H}_{0} = \sum_{m\sigma} \varepsilon^{(0)}_{m\sigma} \hat{n}_{m\sigma},
\end{equation}
where we discarded the hopping amplitude between different orbitals. Here, $\varepsilon^{(0)}_{m\sigma}$ is the on-site energy of orbital $m$ and $\hat{n}_{m\sigma} \left( \equiv \hat{c}^{\dagger}_{m\sigma} \hat{c}_{m\sigma}\right)$ the corresponding particle number operator. The orbitals interact by means of $\hat{H}_{\text{int}} = \hat{H}_{\text{\tiny intra}} + \hat{H}_{\text{\tiny inter}}$, where
\begin{equation}\label{eq:intra_orb_Ham}
\hat{H}_{\text{\tiny intra}} = \frac{U}{2} \sum_{m\sigma}\hat{n}_{m\sigma}\hat{n}_{m\overline{\sigma}},
\end{equation}
and
\begin{equation}\label{eq:inter_orb_Ham}
\hat{H}_{\text{\tiny inter}} = \frac{U^{\prime}}{2} \sum_{\substack{m \neq n}} \sum_{\sigma} \hat{n}_{m\sigma} \hat{n}_{n\overline{\sigma}} + \frac{U^{\prime\prime}}{2} \sum_{\substack{m \neq n}} \sum_{\sigma} \hat{n}_{m\sigma} \hat{n}_{n\sigma},
\end{equation}
account for {\em intra-} and {\em inter-orbital} correlation, respectively. The notation $\overline{\sigma}$ denotes the spin quantum number anti-parallel to $\sigma$.
Here, we neglect spin-flip and pair-hopping terms which is justified in the present case in that we consider impurities with mainly one particle. We recall that due to rotational invariance $U^{\prime} = U - 2J$ and $U^{\prime\prime} = U^{\prime} - J$, with $J$ the Hund's coupling.

The impurity site couples to a fermionic bath described by the following Hamiltonian
\begin{equation}
H_{\text{bath}} = \sum_{mk\sigma} \varepsilon_{mk} \hat{f}^{\dagger}_{mk\sigma} \hat{f}_{mk\sigma},
\end{equation}
where $k$ labels the infinite energy levels of the bath.
Here, we restrict to diagonal coupling between the orbitals on the impurity and the bath,
\begin{equation}
\hat{H}_{\text{i-b}} = \sum_{mk\sigma} \left[ V_{mk} c^{\dagger}_{mk\sigma} f_{mk\sigma} + \text{H.c.} \right],
\end{equation}
which is the case in all the situations we consider in the following. The bath can be completely characterized by the hybridization function, which we introduce below in Eq.~\eqref{eq:orb_dep_hybridization}, which again is diagonal in orbital indices.

\section{Mixed-configuration approximation}\label{sec:method}

To tackle realistic {\em ab initio} problems beyond equilibrium, we approximate the solution to the multi-orbital impurity in Eq.~\eqref{eq:H_tot} by solving a set of independent, single-orbital impurity problems. Each orbital is treated separately at fixed electronic configurations of the other(s); we then weight and combine their solutions according to self-consistently determined probabilities associated with the corresponding configurations. This approximation lets us plug any single-impurity solver into a multi-orbital context. Here, we pair MCA with AMEA, an efficient, steady-state solver for single impurities~\cite{ma.ga.22,ga.ma.22,we.lo.23,ma.we.23,ga.we.24,ku.er.24}.

As a first application of this MCA-AMEA we address a two-orbital impurity, where we label the orbitals in the following as 1 and 2.
In our approach, the (impurity) Green's function (GF) of a given orbital (the {\em target} orbital) is expressed as a {\em weighted sum} of the GFs computed from independent single-impurity problems, each identified by fixing the {\em configuration} of the other orbital(s). Consequently, by targeting a given orbital, one solves as many impurity problems as there are inequivalent configurations, i.e. occupation states for that orbital. This procedure must then be repeated for each orbital in the impurity.

For an impurity consisting of \(N_{\text{orbs}} = 2\) orbitals, the inter-orbital Hamiltonian~\eqref{eq:inter_orb_Ham} reads
\begin{equation}\label{eq:2orb-kanamori_int}
\hat{H}_{\text{\tiny inter}} 
= U^{\prime} \bigl(\hat{n}_{1\uparrow} \hat{n}_{2\downarrow} + \hat{n}_{1\downarrow} \hat{n}_{2\uparrow}\bigr)
+ U^{\prime\prime} \bigl(\hat{n}_{1\uparrow} \hat{n}_{2\uparrow} + \hat{n}_{1\downarrow} \hat{n}_{2\downarrow}\bigr).
\end{equation}
When targeting orbital~1, orbital~2 can be in one of four configurations \(2_{\alpha}\) with \( \alpha \in \mathcal{C} \equiv \{\text{e}, \uparrow, \downarrow, \text{d}\}\), where \(\text{e}\) denotes the {\em empty} and \(\text{d}\) the {\em doubly-occupied} state. In such a fixed configuration, \(\hat{n}_{2\sigma}\) gets replaced by a scalar \(n_{2\sigma}\) that can take the values 0 or 1, depending on its occupancy. 

With this replacement, the Hamiltonian~\eqref{eq:2orb-kanamori_int} will be referred to as \(\hat{\tilde{H}}_{\text{\tiny inter}}(2_{\alpha})\) and becomes a single-particle operator, dependent on the configuration $\alpha$ of orbital 2, which can be added to the non-interacting part of the impurity Hamiltonian~\eqref{eq:nonint_Hamiltonian} for orbital $1$.

If orbital $2$ is empty, i.e. \( 2 \equiv 2_{\text{e}} \), then the on-site energy of orbital~1 is given by
\begin{equation}
\varepsilon_{1\sigma}(2_{\text{e}}) = \varepsilon^{(0)}_{1\sigma},
\end{equation}
where \(\varepsilon^{(0)}_{1\sigma}\) is the on-site energy of the (isolated) orbital~1.\footnote{With the notation \( \varepsilon^{(0)}_{m\sigma} \) we allow for spin-dependent on-site energies for the isolated orbitals. However, in the following we drop the spin index as we always consider them as spin-independent.}

If orbital $2$ is singly occupied by an electron of spin $\sigma^{\prime}$, the on-site energy of orbital~1 clearly becomes spin-dependent
\begin{equation}
\varepsilon_{1\sigma}(2_{\sigma^{\prime}}) = \varepsilon^{(0)}_{1\sigma} + U^{\prime} \delta_{\sigma\overline{\sigma^{\prime}}} + U^{\prime\prime} \delta_{\sigma\sigma^{\prime}}.
\end{equation}
Finally, when  orbital~2 is doubly-occupied (\(2 \equiv 2_{\text{d}}\)) the on-site energy of orbital~1 reads
\begin{equation}
\varepsilon_{1\sigma}(2_{\text{d}}) = \varepsilon^{(0)}_{1\sigma} + U^{\prime} + U^{\prime\prime}.
\end{equation}
To compute the target’s on-site energy for a given configuration, one then adds $U'$ to the isolated orbital term for each electron, the spin of which is anti-parallel to the target’s, and $U''$ for each electron having parallel spin. 

Similarly, for each configuration of orbital~1, we then target orbital~2 and solve the corresponding impurity problem. 

In general, once the configuration $\alpha$ of the other orbital $n$ has been fixed, the impurity problem for the target orbital $m$ is fully defined by $\varepsilon_{m\sigma}(n_{\alpha})$ and $U$. The corresponding configuration-dependent GFs $\kel{G}_{m\sigma}\{n_{\alpha}\}$ can then be evaluated by using the {\em single-impurity} solver of choice. 
Clearly, one can exploit symmetries in this process, for example $\kel{G}_{1\uparrow}\{2_{\downarrow}\} \;=\; \kel{G}_{1\downarrow}\{2_{\uparrow}\}$, whenever possible, to simplify the computations.

As single-impurity solver for the equilibrium and steady-state case we use AMEA with the configuration interaction implementation. 
This approach offers both good accuracy as a single-impurity solver for nonequilibrium steady states and computational efficiency, making it well-suited for DMFT applications, where repeated impurity solutions are required.
For further details of the solver, which we will not repeat here, we refer the reader to  Refs.~\cite{ar.kn.13,do.nu.14, we.lo.23}.

We still need to determine the weights, with which the individual GFs contribute to the multi-orbital GF.  The {\em conditional} probabilities that the target orbital occupies a particular state, given the fixed configuration of the other orbital(s), can be calculated also from the the impurity problem,
\begin{align}\label{eq:conditional_probs_2orb}
\begin{split}
\overline{P}(1_{\text{d}} \mid 2_{\alpha}) & = \left< \hat{n}_{1\text{d}} \right>_{2_\alpha}, \\
\overline{P}(1_{\uparrow} \mid 2_{\alpha}) & = \left< \hat{n}_{1\uparrow} \right>_{2_\alpha} - \left< \hat{n}_{1\text{d}} \right>_{2_\alpha}, \\
\overline{P}(1_{\downarrow} \mid 2_{\alpha}) & =\left< \hat{n}_{1\downarrow}\right>_{2_\alpha} - \left< \hat{n}_{1\text{d}} \right>_{2_\alpha}, \\
\overline{P}(1_{\text{e}} \mid 2_{\alpha}) & = 1 - \sum_{\beta \in \mathcal{C} \setminus \text{e}} \overline{P}(1_{\beta} \mid 2_{\alpha}) \\
& = 1 - \left( \left< \hat{n}_{1\uparrow} \right>_{2_\alpha} + \left< \hat{n}_{1\downarrow} \right>_{2_\alpha} \right) +  \left< \hat{n}_{1\text{d}} \right>_{2_\alpha},
\end{split}
\end{align}
where $\left< \cdots \right>_{2_\alpha}$ denotes the expectation value of an operator of orbital $1$ when orbital $2$ is in configuration $\alpha$ and \(\hat{n}_{1d} \equiv \hat{n}_{1\uparrow} \hat{n}_{1\downarrow}\) is the double occupation operator for orbital~1. Clearly, by construction the conditional probabilities satisfy
\begin{equation}\label{eq:column_stochastic_matrix}
\sum_{\alpha \in \mathcal{C}} \overline{P}(1_{\alpha} \mid 2_{\beta}) = 1.
\end{equation}
Similar definitions hold when orbital~2 is the target and the configuration of orbital~1 is fixed.

The {\em marginal} probabilities defining the occupation of the orbitals are then obtained by calculating
\begin{align}\label{eq:total_probs_2orb}
\begin{split}
P(1_{\alpha}) & = \sum_{\beta \in \mathcal{C}} \overline{P}(1_{\alpha} \mid 2_{\beta})P(2_{\beta}), \\ 
P(2_{\alpha}) & = \sum_{\beta \in \mathcal{C}} \overline{P}(2_{\alpha} \mid 1_{\beta})P(1_{\beta}).
\end{split}
\end{align}
With the probabilities in~\eqref{eq:total_probs_2orb} we then calculate the (orbital) impurity GFs as a weighted sum
\begin{align}\label{eq:imp_GFs_2orb}
\begin{split}
\kel{G}_{1\sigma}(\omega) & = \sum_{\alpha \in \mathcal{C}} P(2_{\alpha}) \ \kel{G}_{1\sigma}\lbrace 2_{\alpha} \rbrace (\omega), \\ 
\kel{G}_{2\sigma}(\omega) & = \sum_{\alpha \in \mathcal{C}} P(1_{\alpha}) \ \kel{G}_{2\sigma} \lbrace 1_{\alpha} \rbrace (\omega).
\end{split}
\end{align}
The {\em configuration-dependent} impurity GFs $\kel{G}_{m\sigma}\lbrace n_{\alpha} \rbrace$ in Eqs.~\eqref{eq:imp_GFs_2orb} satisfy the so-called Keldysh structure~\cite{ma.ga.22,ma.we.23}
\begin{equation}\label{eq:keldysh_structure}
\kel{G} = 
\begin{pmatrix}
G^{\text{R}} & G^{\text{K}} \\
0 & G^{\text{\tiny A}}
\end{pmatrix},
\end{equation}
where \( G^{\text{R}} \), \( G^{\text{K}} \) and \( G^{\text{\tiny A}} = \left[ G^{\text{R}} \right]^{\dagger} \) are the {\em retarded}, {\em Keldysh} and {\em advanced} components, respectively. In principle each of the entries in the Keldysh structure~\eqref{eq:keldysh_structure} can be a matrix, for example in spin indices.

The solution to~\eqref{eq:total_probs_2orb} can be found by recasting the equations into
\begin{align}\label{eq:eig_problem}
\begin{split}
\vec{x} & = M_{1} M_{2} \ \vec{x}, \\
\vec{y} & = M_{2} \ \vec{x},
\end{split}
\end{align}
where the vectors $x_{\alpha}$ and $y_{\alpha}$ have components $P(1_{\alpha})$ and $P(2_{\alpha})$, respectively, and the components of the matrices $M_i$ ($i=1, 2$) are $\overline{P}(1_{\alpha} \mid 2_{\beta})$ and $\overline{P}(2_{\alpha} \mid 1_{\beta})$. 

The solution of the first equation in~\eqref{eq:eig_problem} is given by the eigenvector $\vec{x}$ of $M = M_1 M_2$ corresponding to the eigenvalue $\lambda = 1$. The existence of such an eigenvalue is guaranteed by the column stochastic~\cite{me.tw.93} nature of the conditional probability matrices $M_1$ and $M_2$ as in~\eqref{eq:column_stochastic_matrix}. The eigenvector $\vec{x}$ is then used to find the vector $\vec{y}$ with the second equation in~\eqref{eq:eig_problem}.

In most cases the eigenvector is unique, but in some applications degeneracies might occur.

\section{Results}\label{sec:results}

This section presents our results and is organized as follows. We first benchmark our MCA-AMEA against QMC for multi-orbital equilibrium impurity models, both degenerate and nondegenerate.\footnote{In Appendix~\ref{sec:single_imp}  we also briefly discuss the case of a single impurity at and away from {\em half-filling} to verify the reliability of the AMEA impurity solver independently of our MCA scheme. The results of similar calculations have been already presented in Refs.~\cite{do.nu.14,we.lo.23}.}
Inspired by the ultra-thin SrVO$_{3}$ layer of Ref.~\cite{ba.as.16}, we then apply the method to a toy model of the layered system consisting of two orbitals, performing equilibrium and nonequilibrium DMFT calculations.

For the benchmark of the equilibrium impurity models, we choose {\em box-shaped} orbital hybridization functions,
\begin{align}\label{eq:orb_dep_hybridization}
\begin{split}
\text{Im}\left[ \Delta^{\text{R}}_{m} (\omega) \right] & = - \zeta \left[1 - f_{\text{\tiny FD}}(\omega + D_{m}, T_{\text{fict}}) \right] \\
& \times f_{\text{\tiny FD}}(\omega - D_{m}, T_{\text{fict}}), \\
\text{Im}\left[ \Delta^{\text{K}}_{m} (\omega) \right] & = 2\,\text{Im}\left[ \Delta^{\text{R}}_{m} (\omega) \right] \left[1 - 2f_{\text{\tiny FD}}(\omega - \mu, T) \right],
\end{split}
\end{align}
where the Keldysh component is obtained by the fluctuation-dissipation theorem.~\footnote{The real part of $\Delta^{\text{R}}$ is computed via Kramers-Kronig relations while $\text{Re}\,\left[ \Delta^{\text{K}} \right]$ vanishes throughout the whole energy axis.} Here, 
\[
f_{\text{\tiny FD}}(\omega-\mu,T)=\left[\ee^{(\omega-\mu)/T}+1\right]^{-1}
\]
is the Fermi-Dirac distribution. $D_{m}$ is the half-bandwidth, and $\zeta$ is an orbital-independent energy scaling parameter accounting for the coupling between the impurity and the bath. The parameter $T_{\text{fict}}$ is introduced to smooth the edges of $\text{Im}\left[\Delta^{\text{R}}_{m}(\omega)\right]$. A couple of important observables for our analysis are the (orbital- and spin-resolved) spectrum $A_{m\sigma}(\omega) = -\pi^{-1} \text{Im}\left[ G^{\text{R}}_{m\sigma}(\omega) \right]$ and distribution function 
\[
F_{m\sigma}(\omega) = \left\{ 1 - \text{Im}\left[G^{\text{K}}_{m\sigma}(\omega)\right]/2\text{Im}\left[ G^{\text{R}}_{m\sigma}(\omega) \right] \right\}/2.
\]
Our units are such that \( e = \hbar = k_{\text{\tiny B}} = 1\).

\subsection{Benchmarks - Two-orbital impurity models at equilibrium}

Given that QMC is accurate at equilibrium, we hereby benchmark our MCA-AMEA against it for selected multi-orbital equilibrium impurity models. As QMC solver, we use continuous time quantum Monte Carlo in the hybridisation expansion~\cite{we.co.06} as implemented in the TRIQS/CTHYB package~\cite{pa.fe.15,triqscthyb}.

\subsubsection{Degenerate two-orbital impurity}

We start our investigation with the case of two degenerate orbitals, with $D_{1} = D_{2} = D$ and $\varepsilon^{(0)}_{1} = \varepsilon^{(0)}_{2} = \varepsilon$. All relevant parameters for this setup are listed in Table~\ref{tab:int_same_eps_same_D}.

\begin{figure}[t]
\includegraphics[width=\linewidth]{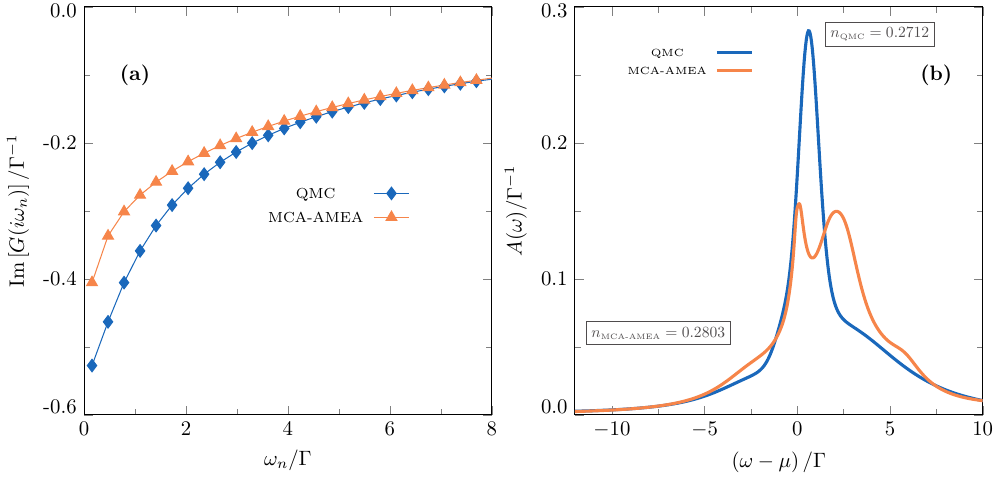}
\caption{(a) Imaginary part of the Matsubara GFs (per spin) for the degenerate two-orbital impurity model at equilibrium obtained with QMC and MCA-AMEA. (b) Corresponding spectra to (a), for QMC obtained from Maximum Entropy~\cite{PhysRevB.96.155128}. The orbital occupations (on the plot) are slightly overestimated by MCA-AMEA. All relevant parameters are listen in Table~\ref{tab:int_same_eps_same_D}.}
\label{fig:GFs_int_same_eps_same_D}
\end{figure}

\begin{table}[h]
    \centering
    \begin{tabular}{c@{\hspace{15pt}} c@{\hspace{15pt}} c@{\hspace{15pt}} c@{\hspace{15pt}}c@{\hspace{15pt}} c@{\hspace{15pt}} c}
        \toprule
        $U$ & $J$ & $\varepsilon$ & $T$ & $D$ & $\mu$ & $\zeta$ \\
        \midrule
        5.5 & 0.75 & 0 & 0.05 & 15 & 2 & 1 \\
        \bottomrule
    \end{tabular}
    \caption{Parameters for the degenerate two-orbital impurity. $D$ denotes the half-bandwidth of $\text{Im}\left[ \Delta^{\text{R}}(\omega)\right]$ and $J$ the Hund's coupling in the inter-orbital part of the Kanamori Hamiltonian. All parameters are given in units of $\Gamma \equiv - \text{Im}\left[ \Delta^{\text{R}}(\omega = 0)\right] / \zeta$, see Eq.~\eqref{eq:orb_dep_hybridization}. The position of the chemical potential produces a slightly more than {\em quarter-filled} impurity. In the rotation-invariant setup $U^{\prime} = U - 2J$ and $U^{\prime\prime} = U^{\prime} - J$.}
    \label{tab:int_same_eps_same_D}
\end{table}
When orbitals are degenerate, MCA-AMEA and QMC give orbital occupations which agree closely.\footnote{See caption of Fig.~\ref{fig:GFs_int_same_eps_same_D} for the exact values.} On the other hand, even though the Matsubara GFs are in reasonable qualitative agreement between the two approaches, as it can be observed from the blue and orange curves in Fig.~\ref{fig:GFs_int_same_eps_same_D}(a), the low-frequency behavior of the QMC solution shows a more pronounced metallicity as evidenced by the larger (absolute) value of $\text{Im}\left[G(i\omega_n)\right]$ at the Fermi level with respect to MCA-AMEA.

The spectra shown in Fig.~\ref{fig:GFs_int_same_eps_same_D}(b) show that for energies below and up to the chemical potential the spectral weight distribution within the two approaches agrees quite well. On the other hand, above the Fermi level they differ appreciably, as our approach shifts a significant portion of the quasi-particle weight away from the Fermi energy. The discrepancy between QMC and MCA-AMEA is most likely to be attributed to the intrinsic limitations of mean-field–like methods\footnote{We point out that our MCA scheme cannot be regarded as a pure mean-field approach in that intra-orbital interactions are treated exactly.} in capturing strong quantum fluctuations in (near-)degenerate systems, a well-known shortcoming discussed extensively in Refs.~\cite{sachdev,al.si.10,balents.10,giamarchi.03,le.na.06}. In future work, our approach needs to be extended in order to cope with such degeneracies.

\subsubsection{Non-degenerate two-orbital impurity}\label{sec:non-deg_two-orb_imp}

We now illustrate how the MCA-AMEA performs for a non-degenerate two-orbital impurity close to {\em quarter-filling}. To lift the orbital degeneracy we choose \(\varepsilon^{(0)}_{1} = 2\Gamma = -\varepsilon^{(0)}_{2}\) while the bath stays the same, i.e. \(D_{1} = D_{2} = D\). In addition, the chemical potential is chosen as the midpoint of the orbital on-site energies. All relevant parameters for this setup are listed in Table~\ref{tab:int_large_crys_field_same_D}.

\begin{table}[b]
    \centering
    \begin{tabular}{c@{\hspace{15pt}} c@{\hspace{15pt}} c@{\hspace{15pt}} c@{\hspace{15pt}} c@{\hspace{15pt}}c@{\hspace{15pt}} c@{\hspace{15pt}} c}
        \toprule
        $U$ & $J$ & $\varepsilon_{1}$ & $\varepsilon_{2}$ & $T$ & $D$ & $\mu$ & $\zeta$ \\
        \midrule
        5.5 & 0.75 & 2 & -2 & 0.05 & 15 & 0 & 1 \\
        \bottomrule
    \end{tabular}
    \caption{The parameters for the non-degenerate two-orbital impurity with on-site energy difference \( \delta\varepsilon \equiv \varepsilon_1 - \varepsilon_2 = 4 \Gamma\). The half-bandwidth of $\text{Im}\left[ \Delta^{\text{R}}(\omega)\right]$ is denoted by $D$ and is the same for the two orbitals. All parameters are given in units of $\Gamma \equiv - \text{Im}\left[ \Delta^{\text{R}}(\omega = 0)\right] / \zeta$, see Eq.~\eqref{eq:orb_dep_hybridization}. For rotational-invariant systems $U^{\prime} = U - 2J$ and $U^{\prime\prime} = U^{\prime} - J$.}
    \label{tab:int_large_crys_field_same_D}
\end{table}

\begin{figure}[t]
\includegraphics[width=\linewidth]{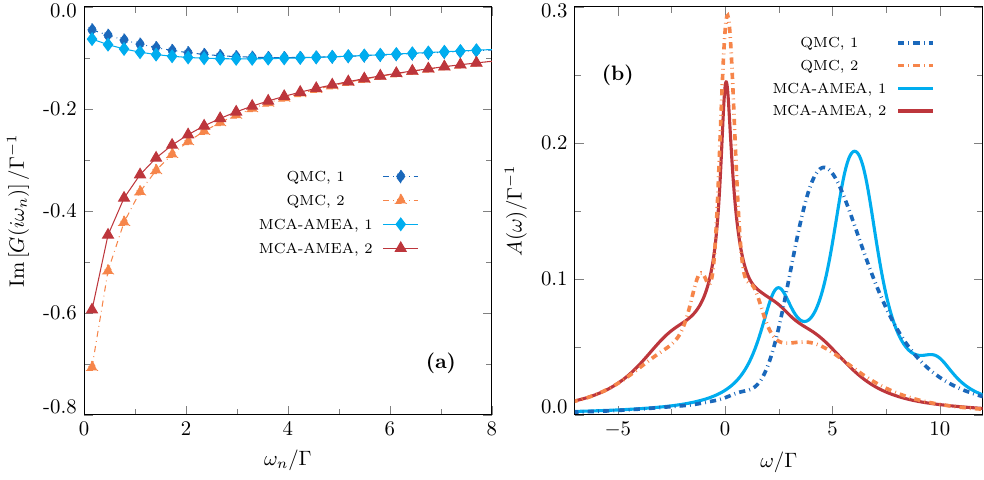}
\caption{(a) Imaginary part of the Matsubara GFs (per spin) for the non-degenerate two-orbital impurity close to {\em quarter-filling} with on-site energy difference \( \delta\varepsilon \equiv \varepsilon^{(0)}_1 - \varepsilon^{(0)}_2=4\Gamma \). The orbital occupations are $n_{\text{\tiny QMC},1} = 0.0499$, $n_{\text{\tiny QMC},2} = 0.4253$ and $n_{\text{\tiny MCA-AMEA},1} = 0.0568$, $n_{\text{\tiny MCA-AMEA},2} = 0.4142$ show a consistent charge polarization for both methods. (b) Corresponding spectra to (a) for both QMC (from Maximum Entropy~\cite{PhysRevB.96.155128}) and MCA-AMEA. The chemical potential is set to $\mu = 0$ while all relevant parameters are listed in Table~\ref{tab:int_large_crys_field_same_D}.}
\label{fig:large_crys_field}
\end{figure}

Given the large difference in the orbital on-site energies, \( \delta\varepsilon \equiv \varepsilon^{(0)}_1 - \varepsilon^{(0)}_2 = 4 \Gamma\), the system shows a strong {\em charge polarization} in favor of orbital~2 in both QMC and MCA-AMEA. This is evident in the Matsubara domain, see Fig.~\ref{fig:large_crys_field}(a), where the imaginary part of GF of orbital~1 clearly extrapolates toward zero, whereas orbital~2 exhibits {\em metallic} behavior, seen from the presence of significant spectral weight at the Fermi level. In real frequency, then, orbital~1 has nearly all its spectral weight above the Fermi level, while orbital~2 features a strong peak at the chemical potential, as shown in Fig.~\ref{fig:large_crys_field}(b). 
We point out the very good agreement between the GFs with the two methods in both Matsubara and real frequency domain. Also the orbital occupations in the two approaches agree very well, see the caption of Fig.~\ref{fig:large_crys_field} for the exact values. This shows that our approximation performs best for impurity setups with non-degenerate orbitals.

\begin{figure*}[t]
    \centering
    \includegraphics[width=\linewidth]{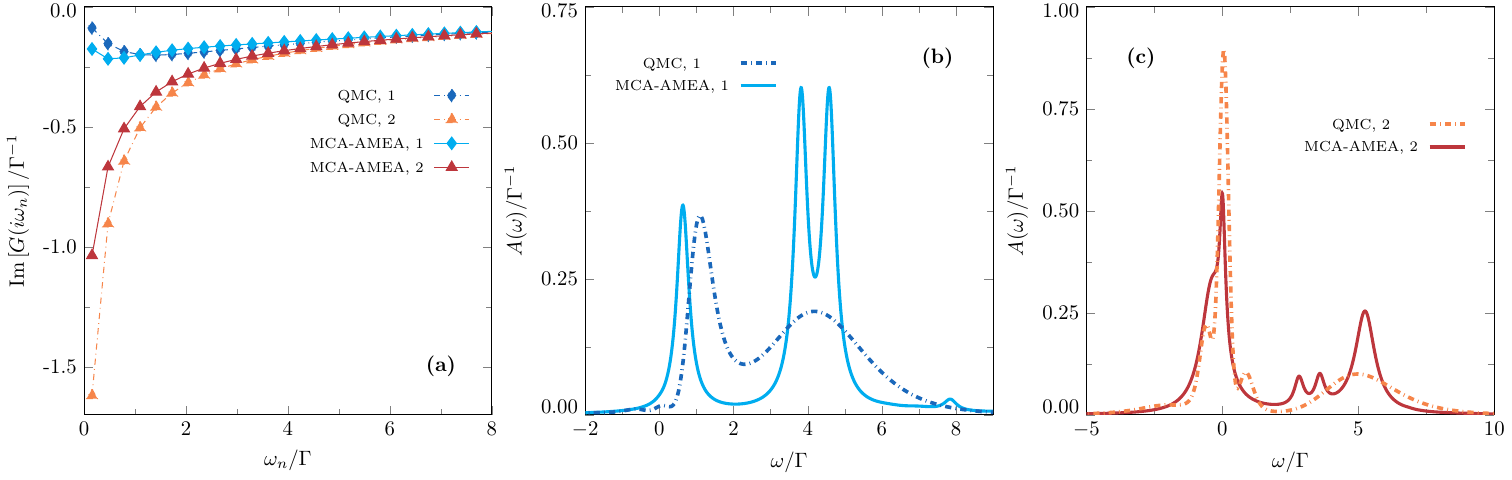}
    \caption{(a) Imaginary part of the Matsubara GFs for a two-orbital impurity with on-site energy difference $\delta\varepsilon = \Gamma$ obtained with QMC and MCA-AMEA. The orbital occupations (per spin) are slightly overestimated by MCA-AMEA: $n_{\mathrm{QMC},1} = 0.026$, $n_{\mathrm{QMC},2} = 0.3516$ and $n_{\mathrm{MCA-AMEA},1} = 0.033$, $n_{\mathrm{MCA-AMEA},2} = 0.3763$. However, the overall agreement between the two solvers stays good. (b) and (c) show the corresponding spectra of orbitals~1 and~2, respectively, for both solvers. The chemical potential is set to $\mu = 0$ and all relevant parameters are listed in Table~\ref{tab:int_small_crys_field_same_D}.}
    \label{fig:small_crys_field}
\end{figure*}

However, the on-site energy separation considered here, see Table~\ref{tab:int_large_crys_field_same_D}, is too large to be physically relevant for realistic systems. In the following, we then investigate a setup with a smaller orbital on-site energy difference, $\delta\varepsilon = \Gamma$. 
Table~\ref{tab:int_small_crys_field_same_D} lists the parameters for such a two-orbital system with notably smaller crystal field splitting. 
Given the reduced $\delta\varepsilon$, we also decrease the coupling strength $\zeta$ between the bath and the impurity in order to avoid overshadowing important features in the resulting GFs. 

In this setup, MCA-AMEA still reproduces the QMC results with reasonable accuracy, as can be seen from the imaginary parts of the Matsubara GFs in Fig.~\ref{fig:small_crys_field}(a) and the spectra in Figs.~\ref{fig:small_crys_field}(b) and (c). The orbital occupations deviate by only a few percentage points: the exact values can be found in the caption of Fig.~\ref{fig:small_crys_field}. Nonetheless, the qualitative shape of the spectra is quite satisfactory throughout the whole energy range. 
\begin{table}[t]
    \centering
    \begin{tabular}{c@{\hspace{15pt}} c@{\hspace{15pt}} c@{\hspace{15pt}} c@{\hspace{15pt}} c@{\hspace{15pt}}c@{\hspace{15pt}} c@{\hspace{15pt}} c}
        \toprule
        $U$ & $J$ & $\varepsilon_{1}$ & $\varepsilon_{2}$ & $T$ & $D$ & $\mu$ & $\zeta$ \\
        \midrule
        5.5 & 0.75 & 0.5 & -0.5 & 0.05 & 15 & 0 & 0.2 \\
        \bottomrule
    \end{tabular}
    \caption{Parameters for the two-orbital non-degenerate impurity with on-site energy difference $\delta\varepsilon=\Gamma$. The half-bandwidth of the bath, $D$, is the same for both orbitals. All parameters are in units of $\Gamma \equiv -\mathrm{Im}[\Delta^{\mathrm{R}}(\omega=0)]/\zeta$, see Eq.~\eqref{eq:orb_dep_hybridization}. For rotational-invariant systems, $U^{\prime} = U - 2J$ and $U^{\prime\prime} = U' - J$.}
    \label{tab:int_small_crys_field_same_D}
\end{table}
When paired with the previous findings, these results confirm that the MCA-AMEA performs best when the orbitals are nondegenerate. However, when the on-site energy difference is reduced, the overall agreement between MCA-AMEA and QMC is slightly worse than the case with larger $\delta\varepsilon$ discussed in Fig.~\ref{fig:large_crys_field}.

\subsection{Two-orbital lattice system}

Motivated by the findings of Ref.~\cite{ba.as.16}, we test our method's performance for a realistic two-orbital lattice model inspired by the ultra-thin SrVO$_3$ film.

As discussed in Sec.~\ref{sec:intro}, growing SrVO$_3$ in very thin films partially lifts the degeneracy of the \(t_{2g}\) manifold. In the bulk, each of the three \(t_{2g}\) orbitals holds 1/3 of an electron and the compound is metallic. In few-layer systems, however, an insulating phase develops: a crystal-field splitting arises between the two-dimensional (2D) \(d_{xy}\) orbital and the one-dimensional (1D) \(d_{yz}\) and \(d_{zx}\) orbitals. As shown in Ref.~\cite{ba.as.16}, this splitting—combined with the different orbital dimensionalities—is responsible for the charge migration to orbital \(d_{xy}\), resulting in almost complete charge polarization.

Here, we start from a monolayer of SrVO$_3$ as in Ref.~\cite{ba.as.16}. The DFT Hamiltonian has been constructed using the Wien2k (version 14.2)~\cite{Blaha_2020_PAPER}, TRIQS/DFTTools (version 3.3)~\cite{pa.fe.15,Aichhorn2016}, and Wannier90 (version 3.0)~\cite{Pizzi2020} packages. For the sake of simplicity, we drop one of the two remaining degenerate 1D orbitals. Hence we retain only $d_{xy}$ and $d_{yz}$ in the effective Hamiltonian. The chemical potential is set such that we obtain a total occupation of 1.0 electron in these two orbitals. This is a sufficient starting point for our 
prototype
 study here, we postpone the study of the full three-orbital lattice to a later work. Input files for the calculation of the DFT and Wannier Hamiltonians are publicly available from the research data server of TU Graz~\cite{data}.

We apply MCA-AMEA within the DMFT framework to this toy-model lattice and benchmark the results again against QMC.

The DMFT loop proceeds by enforcing that, at self‐consistency, the impurity ($\mathbf{\kel{G}}$) and local ($\mathbf{\kel{G}}_{\text{loc}}$) GFs coincide, 
\begin{equation}
\mathbf{\kel{G}}(\omega)\;\overset{!}{=}\;\mathbf{\kel{G}}_{\text{loc}}(\omega)\,.
\end{equation}
To achieve this, we update the hybridization function as
\begin{equation}
\mathbf{\kel{\Delta}}(\omega)
=\mathbf{\kel{g}}_{0}^{-1}(\omega)
-\mathbf{\kel{G}}_{\mathrm{loc}}^{-1}(\omega)
-\mathbf{\kel{\Sigma}}(\omega)\,,
\end{equation}
where $\mathbf{g}_{0}^{-1,\text{R}}(\omega) = \left( \omega - \boldsymbol{\varepsilon}^{(0)}\right) \mathbb{1}$ and $\mathbf{\kel{\Sigma}}$ is the impurity self-energy (SE). The impurity problem is then solved and the procedure iterated until convergence is reached.\footnote{A boldface character denotes a matrix in both orbital and spin quantum numbers.}
In accordance with Refs.~\cite{zh.wa.15,ba.as.16}, we set the Hubbard interaction and Hund's coupling as $U = 5.5 \, \mathrm{eV}$ and $J = 0.75\, \mathrm{eV}$. From now on, we give all energies in units of eV.

\subsection{Equilibrium: The role of Hund's coupling}

\begin{figure*}[t]
\includegraphics[width=\linewidth]{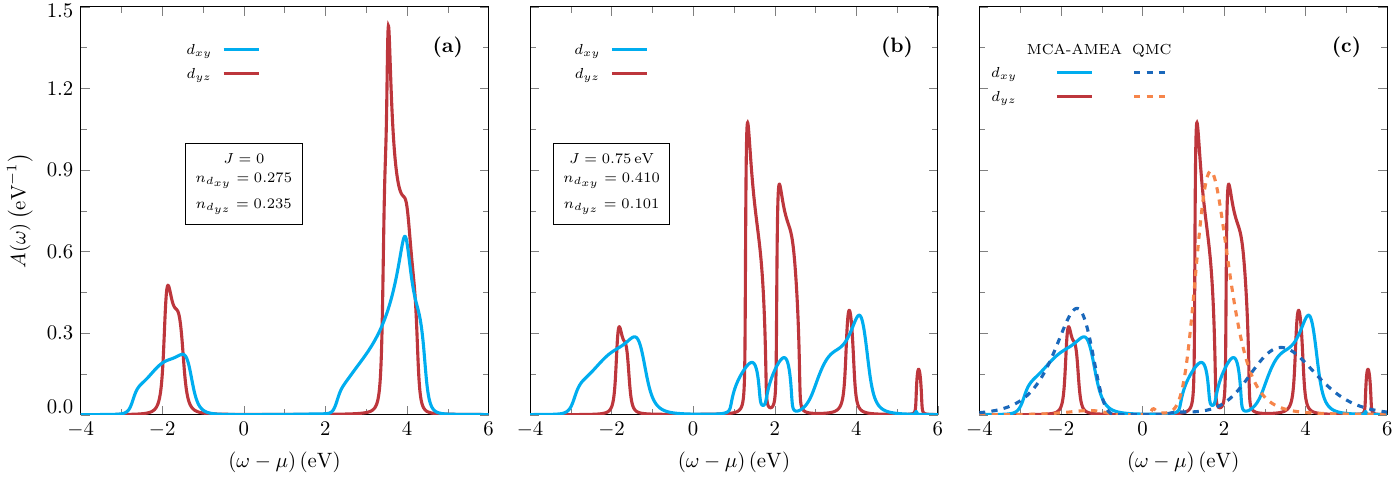}
\caption{(a) Equilibrium DMFT spectra for the $d_{xy}$ and $d_{yz}$ orbitals at $J=0$ obtained with MCA-AMEA at 1/4 filling. The gap is of the order of the Hubbard interaction $U$ and the orbital occupations remain nearly equal, indicating minimal polarization. (b) Same as (a) for $J=0.75\,\mathrm{eV}$. Hund’s coupling reduces the gap to approximately $U - 3J$ and enhances orbital polarization: $d_{xy}$ is occupied about four times more than $d_{yz}$. (c) Comparison between the equilibrium DMFT spectra obtained with MCA-AMEA (already plotted in (b)) and QMC. The orbital occupations per spin in QMC are $n_{d_{xy}}=0.479$ and $n_{d_{yz}}=0.014$. The chemical potential is $\mu = 2 \, \mathrm{eV}$ (inside the gap) while temperature is set to $T = 0.025 \, \mathrm{eV}$.}
\label{fig:2orbs_dmft}
\end{figure*}
In Fig.~\ref{fig:2orbs_dmft}(a) we show the converged equilibrium DMFT orbital spectral functions obtained with MCA-AMEA at \(J=0\). Both \(d_{xy}\) and \(d_{yz}\) exhibit a pronounced gap of order \(U\), while the Hubbard bands of orbital \(d_{xy}\) are roughly twice as large as those of \(d_{yz}\), indicating their different dimensionality. In this setup, the occupations of the two orbitals remain nearly equal, pointing at negligible polarization.

When Hund’s coupling is set to $J = 0.75\, \mathrm{eV}$, see Fig.~\ref{fig:2orbs_dmft}(b), the gap shrinks to roughly $U - 3J$. The main peaks in which the upper Hubbard band splits into are at energies roughly equal to $U - 3J$, $U - 2J$, and $U$. At the same time, the weights of the lower Hubbard bands become dominated by the $d_{xy}$ orbital—its occupancy is now about four times larger than that of $d_{yz}$. This enhancement in the charge polarization can be understood in terms of the partial lifting of the degeneracy of the orbital configurations as $J$ is turned on, which decreases the overall symmetry of the system. To sum up, our mixed‐configuration approximation captures a strong, though partial, charge polarization, with most of the charge shifting into $d_{xy}$ as compared to the setup with $J=0$. Hund’s coupling appears to be an important driver of this effect in this approximate scheme.

Finally, in Fig.~\ref{fig:2orbs_dmft}(c), we compare the equilibrium spectra obtained with MCA-AMEA and QMC. The gap position and width are well reproduced by our method. However, the upper Hubbard bands of both orbitals in MCA-AMEA are split into several peaks that are absent in the QMC result, a feature which is reminiscent of the way we compute the (average) impurity GF, see Eqs.~\eqref{eq:imp_GFs_2orb}. The lower Hubbard band of the \(d_{yz}\) orbital also retains more spectral weight in MCA-AMEA with respect to QMC, consistent with the corresponding orbital occupations (see the caption of Fig.~\ref{fig:2orbs_dmft} for details). Despite these differences, MCA-AMEA captures the key physical features of the system, although less prominently than QMC.

\subsection{Nonequilibrium: Current-voltage characteristics}

Here we discuss the results of the 
prototype
 nonequilibrium setup where the layer is sandwiched between two metallic baths having different chemical potentials. The so-called wide-band limit approximation is employed for the baths' GFs, so that they contribute the (momentum-independent) {\em embedding} SE
\begin{align}\label{eq:bath_SE}
\begin{split}
\Sigma^{\text{R}}_{\text{bath}, \rho}(\omega) & = - \frac{\ii}{2}\gamma_{\rho}, \\
\Sigma^{\text{K}}_{\text{bath}, \rho}(\omega) & = 2\ii \text{Im}\left[ \Sigma^{\text{R}}_{\text{bath}, \rho}(\omega) \right] \left( 1 - 2 f_{\text{\tiny FD}}(\omega - \mu_{\rho}, T) \right),
\end{split}
\end{align}
where \( \rho \in \{ \text{l}, \text{r} \} \) labels the left and right bath and \( \gamma_{\rho}\) is the corresponding dissipative term. By adding the embedding SE of the baths to the electronic SE, the local GF is then obtained as
\begingroup
\small
\begin{equation}\label{eq:GF_lattice}
\kel{\bm{G}}_{\text{loc}}(\omega) = \frac{1}{V} \sum_{\vec{k}} \left[ \kel{\bm{G}}^{-1}_{0,\text{\tiny DFT}}(\vec{k},\omega) - \sum_{\rho} \kel{\bm{\Sigma}}_{\text{bath}, \rho}(\omega) - \kel{\bm{\Sigma}}(\omega) \right]^{-1}.
\end{equation}
\endgroup
The chemical potentials of the baths are shifted symmetrically around the equilibrium value by the application of the bias voltage \( \Phi \equiv \mu_{\text{l}} - \mu_{\text{r}} \). This potential drop between the two ends of the layer gives rise to a steady-state current perpendicular to it and can be computed by means of the Meir-Wingreen formula as in previous work~\cite{ga.ma.22,ga.we.24},
\begin{equation}\label{eq:steady-state_current}
J_{\text{l} \to \text{r}} = \text{Re} \left\{  \sum_{m\sigma} \int_{-\infty}^{\infty} \frac{\dd\omega}{2\pi} J_{m\sigma}(\omega) \right\}.
\end{equation}
The integrand in Eq.~\eqref{eq:steady-state_current}, $J_{m\sigma}(\omega)$, is defined as
\begin{align}\label{eq:current_integrand}
\begin{split}
J_{m\sigma}(\omega) & = \left[ G^{\text{R}}_{\text{loc}}(\omega)\right]_{m\sigma} \left( \Sigma^{\text{K}}_{\text{bath}, \text{l}}(\omega) - \Sigma^{\text{K}}_{\text{bath}, \text{r}}(\omega) \right) \\
& + \left[ G^{\text{K}}_{\text{loc}}(\omega)\right]_{m\sigma} \left( \Sigma^{\text{A}}_{\text{bath}, \text{l}}(\omega) - \Sigma^{\text{A}}_{\text{bath}, \text{r}}(\omega) \right),
\end{split}
\end{align}
where we dropped spin and orbital indices for the various components of the baths SE as they do not depend on them.

\begin{figure*}[t]
\includegraphics[width=\linewidth]{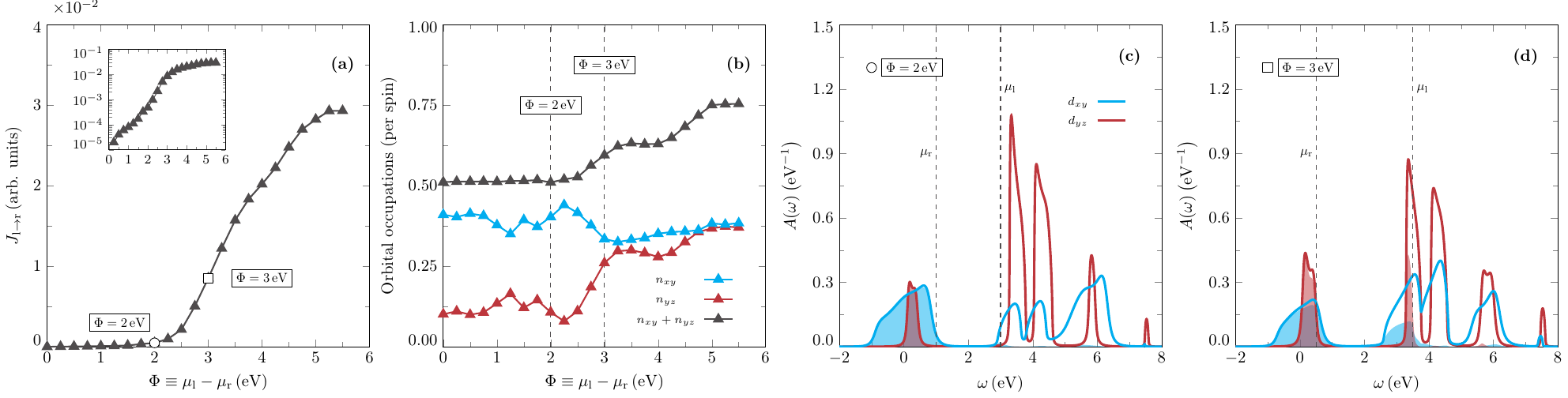}
\caption{(a) Nonequilibrium current density $J_{\text{l}\to\text{r}}$ [Eqs.~\eqref{eq:steady-state_current} and~\eqref{eq:current_integrand}] as function of the applied bias voltage $\Phi$. Inset shows the same plot in a semi-log scale to emphasize the several orders of magnitude spanned by the current. (b) Orbital occupations (per spin) as a function of $\Phi$. For $\Phi \in [2.25, 2.5, \dots, 3]$ a consistent amount of charge is redistributed from $d_{xy}$ to $d_{yz}$. (c) Orbital spectra at $\Phi = 2\, \mathrm{eV}$: the bias voltage is just enough to bridge the gap, giving rise to the first noticable increase in the current, see the corresponding symbol in (a). (d) Same as in (c) at $\Phi=3\,\mathrm{eV}$, well into the metallic regime, see corresponding symbol in (a). Shaded areas in (c) and (d) denote the orbital occupation function $N(\omega) = F(\omega)A(\omega)$, with $F(\omega)$ defined at the beginning of Sec.~\ref{sec:results}. Temperature is set to $T = 0.025 \, \mathrm{eV}$.}
\label{fig:2orbs_neq_dmft}
\end{figure*}

We initialize the nonequilibrium runs from the converged DMFT solution in Fig.~\ref{fig:2orbs_dmft}(b) at $J = 0.75\,\mathrm{eV}$, then apply a symmetric bias to the two chemical potentials around the equilibrium value. As shown in Fig.~\ref{fig:2orbs_neq_dmft}(a), the steady‐state current $J_{\mathrm{l}\to\mathrm{r}}$ remains negligible until the applied voltage $\Phi$ equals the gap energy. Beyond this threshold, $\mu_{\text{l}}$ and $\mu_{\text{r}}$ enter the upper and lower Hubbard bands, respectively, allowing a steady particle flow and causing the current to increase monotonically with $\Phi$. The several orders of magnitude spanned by the current can be better appreciated from the inset in Fig.~\ref{fig:2orbs_neq_dmft}(a). 
Fig.~\ref{fig:2orbs_neq_dmft}(b) shows the orbital occupations (per spin) as a function of $\Phi$. Below $\Phi = 2\,\mathrm{eV}$, the system stays at half‐filling, see also the spectra in Fig.~\ref{fig:2orbs_neq_dmft}(c), despite the presence of oscillations in the individual orbital occupations as $\Phi$ grows to meet its threshold value. When $\Phi$ first matches the gap, a charge rearrangement starts to occur: the overall increase of the total occupation is dominated by the partial discharge of the (initially more occupied) $d_{xy}$ orbital, by virtue of its larger bandwidth, in favor of orbital $d_{yz}$. When $\Phi$ increases further, $\mu_{\text{l}}$ aligns with the upper band of $d_{yz}$ and injects a consistent amount of electrons into the high-energy regions of the 1D orbital, see also Fig.~\ref{fig:2orbs_neq_dmft}(c). For $\Phi \ge 4\,\mathrm{eV}$ the charge rearrangement softens and orbital occupations converge toward the same value.

\section{Conclusions}\label{sec:conclusions}

In this work we introduced a mixed-configuration approximation to address multi-orbital impurity systems in and out of equilibrium. The approach is quite versatile as it can be based on any single-band impurity solver: in this work we deploy it using the AMEA impurity solver for both impurity models and realistic lattice systems in and out of equilibrium.
In the former case, our approximation works best when the orbitals are nondegenerate as it has reduced capabilities to resolve the quantum fluctuations stemming from the high symmetry (degeneracy) of the system. However, also in this case the occupations are rather accurate when compared to QMC, indicating that the differences are most likely related to the unoccupied states of the system, corresponding to regions of the spectrum above the Fermi energy.

In combination with DMFT, our approximation can also be used to study realistic lattice systems in and out of equilibrium conditions. Inspired by previous work~\cite{ba.as.16}, we showed that for a two-orbital layered system at equilibrium our solver predicts charge polarization, even though not as strong as in QMC.

We also addressed a prototype nonequilibrium setup consisting of the aforementioned two-orbital layer sandwiched between metallic contacts at different chemical potentials, creating a constant bias voltage. Starting from a polarized state, we observe charge flowing from the initially more occupied orbital into the other as the bias exceeds the band gap. When the voltage approaches twice the gap, the orbital occupations converge towards each other and the polarization diminishes.

There are still several aspects that must be addressed for this approach to make it applicable for more complex settings in realistic materials modelling. First, extending the procedure described in Eqs.~\eqref{eq:conditional_probs_2orb} to~\eqref{eq:total_probs_2orb} to the case of three (and more) orbitals requires incorporating the joint probability of two orbitals being in a specific configuration. This necessitates accounting for the correlation probabilities between these orbitals. Second, in materials with two or more particles per atom, spin-flip and pair-hopping terms might become relevant. As for now, it seems that extending the present configuration-based approach to capture these effects is nontrivial. In addition, the case of degenerate orbitals must be improved. One possible solution to this aspect is to generalize the method using a "cumulant" expansion, following the ideas in Refs.~\cite{kubo.62,sh.we.80}, in which "connected clusters" involving spin degrees of freedom on different orbitals are explicitly considered. Work in this direction is currently underway.

In summary, the approach presented here represents a first step towards the description of the nonequilibrium behavior of realistic lattice systems directly in the frequency domain. It is worth noting that, while the MCA-AMEA method has moderate computational costs, its main significance lies in enabling the direct study of nonequilibrium steady states in quantum systems, a regime that, to the best of our knowledge, is not easily accessible to quantum Monte Carlo at the present stage (an exception is steady-state inchworm, which is however computationally very expensive). 
In future work, we will extend the MCA scheme to systems with more than two orbitals, e.g., the full SrVO$_3$ layer of Ref.~\cite{ba.as.16}. Furthermore, we plan to explore new geometries and driving conditions. One direction is to apply a static electric field in the plane of the layer rather than perpendicular to it. Another is to study heterostructures composed of multiple parallel layers under constant bias. These extensions will test the versatility of our approach and bring us closer to modeling nonequilibrium phenomena in realistic materials.

\appendix

\section{Single-orbital impurity model at equilibrium}\label{sec:single_imp}

Although the AMEA solver’s performance for the single-orbital impurity model was already assessed (see, for instance, Ref.~\cite{we.lo.23}), we revisit those results to show that any deviation from the equilibrium QMC multi-orbital data discussed in the main text comes from our mixed-configuration approach.

In the following we revisit the standard Anderson impurity model (AIM) at and away from half-filling.

\subsection{Anderson impurity model at half-filling}

\begin{figure}[t]
\includegraphics[width=\linewidth]{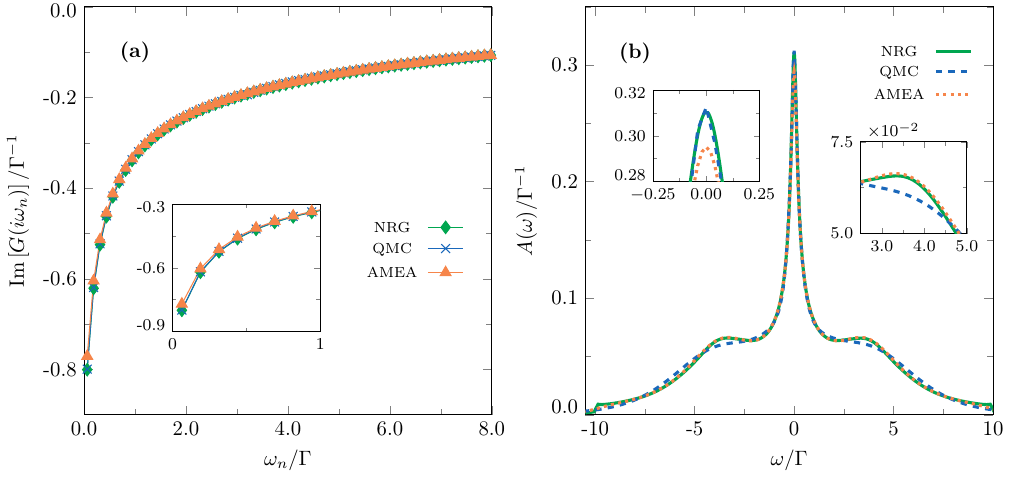}
\caption{(a) Imaginary part of the Matsubara GF of the AIM at half-filling obtained with NRG, QMC and AMEA. (b) Spectra corresponding to (a). The QMC result has been obtained with the MEM of the TRIQS package~\cite{PhysRevB.96.155128}. While the method is very accurate in reproducing features close to the Fermi level ($\mu = 0$), it proves less reliable in recovering the exact shape of the lower and upper Hubbard bands at $\omega \approx \pm U/2$, see insets. All relevant parameters are listed in Table~\ref{tab:AIM_ph_sym}.}
\label{fig:aim_gfs_ph_sym}
\end{figure}

Here we present the results of the AIM at half-filling. All relevant parameters can be found in Table~\ref{tab:AIM_ph_sym}.
\begin{table}[b]
    \centering
    \begin{tabular}{c@{\hspace{15pt}} c@{\hspace{15pt}} c@{\hspace{15pt}} c@{\hspace{15pt}} c@{\hspace{15pt}} c}
        \toprule
        $U$ & $\varepsilon$ & $T$ & $D$ & $\mu$ & $\zeta$\\
        \midrule
        6 & -3 & 0.02 & 10 & 0 & 1 \\
        \bottomrule
    \end{tabular}
    \caption{Parameters of the AIM in the particle-hole symmetric case. $D$ is the half-bandwidth of $\text{Im}\left[ \Delta^{\text{R}}(\omega)\right]$ and $\Gamma \equiv - \text{Im}\left[ \Delta^{\text{R}}(\omega = 0)\right] / \zeta$ sets the unit of energy, see Eq.~\eqref{eq:orb_dep_hybridization}.}
    \label{tab:AIM_ph_sym}
\end{table}
Figure~\ref{fig:aim_gfs_ph_sym}(a) shows the Matsubara GFs obtained with NRG, QMC, and AMEA. The NRG and QMC curves overlap almost exactly, while AMEA shows slight deviations at low energies: overall agreement, however, remains excellent.

Figure~\ref{fig:aim_gfs_ph_sym}(b) presents the corresponding spectra. Even at the low temperature used here ($T = 0.02\,\Gamma$), the quasi‐particle peak at $\omega = 0$ (inset) from AMEA matches the QMC and NRG results closely. The upper and lower Hubbard bands at $\omega \approx \pm U/2$ are also well captured by AMEA. In QMC, these bands appear less distinct due to the known limitations of analytic continuation~\cite{gu.ja.91,be.go.00}, which, however, do not compromise the results of the maximum‐entropy method (MEM) near the Fermi level.

\subsection{Anderson impurity model away from half-filling}

We continue our investigation of the AIM away from half-filling. All relevant parameters are listed in Table~\ref{tab:AIM_nonph_sym}.

\begin{table}[b]
    \centering
    \begin{tabular}{c@{\hspace{15pt}} c@{\hspace{15pt}} c@{\hspace{15pt}} c@{\hspace{15pt}} c@{\hspace{15pt}} c}
        \toprule
        $U$ & $\varepsilon$ & $T$ & $D$ & $\mu$ & $\zeta$\\
        \midrule
        5.5 & 0 & 0.05 & 15 & 0 & 1 \\
        \bottomrule
    \end{tabular}
    \caption{Parameters of the AIM away from half-filling. Here $D$ is the half-bandwidth of $\text{Im}\left[ \Delta^{\text{R}}(\omega)\right]$ and $\Gamma \equiv - \text{Im}\left[ \Delta^{\text{R}}(\omega = 0)\right] / \zeta$ sets the unit of energy, see Eq.~\eqref{eq:orb_dep_hybridization}.}
    \label{tab:AIM_nonph_sym}
\end{table}

\begin{figure}[t]
\includegraphics[width=\linewidth]{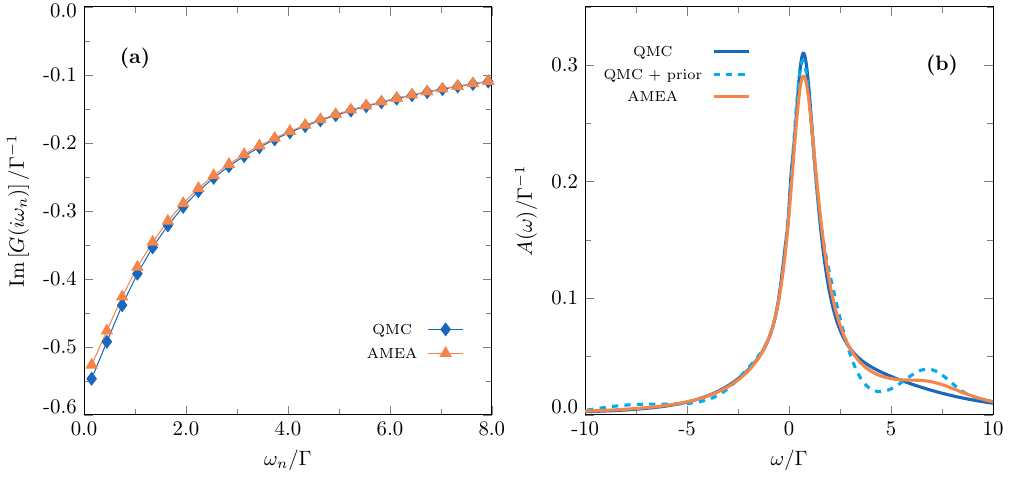}
\caption{(a) Imaginary part of the Matsubara GF of the AIM away from half-filling obtained with the AMEA and QMC. The orbital occupations (per spin) are $n_{\text{\tiny AMEA}}=0.2612$ and $n_{\text{\tiny QMC}}=0.2637$. (b) Spectra corresponding to panel (a). The MEM accurately reproduces the spectral weight near the chemical potential but misses the side peak of the upper band at $\omega \approx 7\Gamma$. By preconditioning the MEM with a prior that includes a peak in this region one recovers the feature, albeit in a non-unique manner. The chemical potential is set to $\mu=0$ and all other parameters are listed in Table~\ref{tab:AIM_nonph_sym}.}
\label{fig:GFs_detached_nonph_sym}
\end{figure}

The AMEA impurity solver proves reliable even in this setup, as evidenced by a direct comparison with the QMC results in Figs.~\ref{fig:GFs_detached_nonph_sym}(a) and (b). Figure~\ref{fig:GFs_detached_nonph_sym}(a) shows very good agreement between the Matsubara GFs obtained with the two solvers, with the orbital occupations (per spin) underestimated by only about $1\%$ in the AMEA results, see the caption of Fig.~\ref{fig:GFs_detached_nonph_sym} for the exact values. Figure~\ref{fig:GFs_detached_nonph_sym}(b) displays the corresponding spectra in both approaches. Here, the agreement is also good, particularly around the Fermi level ($\mu=0$). However, the MEM fails to capture the upper Hubbard band located at $\omega/\Gamma \sim 7$, which is clearly visible in the spectrum obtained with AMEA. It is worth noting that by preconditioning the MEM with a prior that accounts for the satellite peak one could, in principle, reconstruct some spectral weight near the upper band, see again Fig.~\ref{fig:GFs_detached_nonph_sym}(b). Nevertheless, this procedure is highly sensitive to the details and is therefore not very reliable. In summary, we have demonstrated that the AMEA impurity solver agrees quite well with the QMC approach for detached orbitals both at and away from half-filling at equilibrium. 

\begin{acknowledgments}
This research was funded by the Austrian Science Fund (FWF) [Grant DOI:10.55776/P33165], and by NaWi Graz. For the purpose of open access, the author has applied a CC BY public copyright licence to any Author Accepted Manuscript version arising from this submission. Results have been obtained using the A-Cluster at TU Graz as well as the Austrian Scientific Computing (ASC) infrastructure.
\end{acknowledgments}

\section*{Data Availability}
The data that support the findings of this study are openly available in the TU Graz Repository~\cite{data}.

\bibliography{references_database,my_refs}

\begin{thebibliography}{58}%
\makeatletter
\providecommand \@ifxundefined [1]{%
 \@ifx{#1\undefined}
}%
\providecommand \@ifnum [1]{%
 \ifnum #1\expandafter \@firstoftwo
 \else \expandafter \@secondoftwo
 \fi
}%
\providecommand \@ifx [1]{%
 \ifx #1\expandafter \@firstoftwo
 \else \expandafter \@secondoftwo
 \fi
}%
\providecommand \natexlab [1]{#1}%
\providecommand \enquote  [1]{``#1''}%
\providecommand \bibnamefont  [1]{#1}%
\providecommand \bibfnamefont [1]{#1}%
\providecommand \citenamefont [1]{#1}%
\providecommand \href@noop [0]{\@secondoftwo}%
\providecommand \href [0]{\begingroup \@sanitize@url \@href}%
\providecommand \@href[1]{\@@startlink{#1}\@@href}%
\providecommand \@@href[1]{\endgroup#1\@@endlink}%
\providecommand \@sanitize@url [0]{\catcode `\\12\catcode `\$12\catcode
  `\&12\catcode `\#12\catcode `\^12\catcode `\_12\catcode `\%12\relax}%
\providecommand \@@startlink[1]{}%
\providecommand \@@endlink[0]{}%
\providecommand \url  [0]{\begingroup\@sanitize@url \@url }%
\providecommand \@url [1]{\endgroup\@href {#1}{\urlprefix }}%
\providecommand \urlprefix  [0]{URL }%
\providecommand \Eprint [0]{\href }%
\providecommand \doibase [0]{https://doi.org/}%
\providecommand \selectlanguage [0]{\@gobble}%
\providecommand \bibinfo  [0]{\@secondoftwo}%
\providecommand \bibfield  [0]{\@secondoftwo}%
\providecommand \translation [1]{[#1]}%
\providecommand \BibitemOpen [0]{}%
\providecommand \bibitemStop [0]{}%
\providecommand \bibitemNoStop [0]{.\EOS\space}%
\providecommand \EOS [0]{\spacefactor3000\relax}%
\providecommand \BibitemShut  [1]{\csname bibitem#1\endcsname}%
\let\auto@bib@innerbib\@empty
\bibitem [{\citenamefont {Janod}\ \emph {et~al.}(2015)\citenamefont {Janod},
  \citenamefont {Tranchant}, \citenamefont {Corraze}, \citenamefont
  {Querr{\'e}}, \citenamefont {Stoliar}, \citenamefont {Rozenberg},
  \citenamefont {Cren}, \citenamefont {Roditchev}, \citenamefont {Phuoc},
  \citenamefont {Besland},\ and\ \citenamefont {Cario}}]{ja.tr.15}%
  \BibitemOpen
  \bibfield  {author} {\bibinfo {author} {\bibfnamefont {E.}~\bibnamefont
  {Janod}}, \bibinfo {author} {\bibfnamefont {J.}~\bibnamefont {Tranchant}},
  \bibinfo {author} {\bibfnamefont {B.}~\bibnamefont {Corraze}}, \bibinfo
  {author} {\bibfnamefont {M.}~\bibnamefont {Querr{\'e}}}, \bibinfo {author}
  {\bibfnamefont {P.}~\bibnamefont {Stoliar}}, \bibinfo {author} {\bibfnamefont
  {M.}~\bibnamefont {Rozenberg}}, \bibinfo {author} {\bibfnamefont
  {T.}~\bibnamefont {Cren}}, \bibinfo {author} {\bibfnamefont {D.}~\bibnamefont
  {Roditchev}}, \bibinfo {author} {\bibfnamefont {V.~T.}\ \bibnamefont
  {Phuoc}}, \bibinfo {author} {\bibfnamefont {M.-P.}\ \bibnamefont {Besland}},\
  and\ \bibinfo {author} {\bibfnamefont {L.}~\bibnamefont {Cario}},\ }\bibfield
   {title} {\bibinfo {title} {Resistive switching in mott insulators and
  correlated systems},\ }\href {https://doi.org/10.1002/adfm.201500823}
  {\bibfield  {journal} {\bibinfo  {journal} {Advanced Functional Materials}\
  }\textbf {\bibinfo {volume} {25}},\ \bibinfo {pages} {6287} (\bibinfo {year}
  {2015})}\BibitemShut {NoStop}%
\bibitem [{\citenamefont {Metzner}\ and\ \citenamefont
  {Vollhardt}(1989)}]{me.vo.89}%
  \BibitemOpen
  \bibfield  {author} {\bibinfo {author} {\bibfnamefont {W.}~\bibnamefont
  {Metzner}}\ and\ \bibinfo {author} {\bibfnamefont {D.}~\bibnamefont
  {Vollhardt}},\ }\bibfield  {title} {\bibinfo {title} {Correlated lattice
  fermions in $d=\infty$ dimensions},\ }\href
  {https://doi.org/10.1103/PhysRevLett.62.324} {\bibfield  {journal} {\bibinfo
  {journal} {Phys. Rev. Lett.}\ }\textbf {\bibinfo {volume} {62}},\ \bibinfo
  {pages} {324} (\bibinfo {year} {1989})}\BibitemShut {NoStop}%
\bibitem [{\citenamefont {Georges}\ and\ \citenamefont
  {Kotliar}(1992)}]{ge.ko.92}%
  \BibitemOpen
  \bibfield  {author} {\bibinfo {author} {\bibfnamefont {A.}~\bibnamefont
  {Georges}}\ and\ \bibinfo {author} {\bibfnamefont {G.}~\bibnamefont
  {Kotliar}},\ }\bibfield  {title} {\bibinfo {title} {Hubbard model in infinite
  dimensions},\ }\href@noop {} {\bibfield  {journal} {\bibinfo  {journal}
  {Phys. Rev. B}\ }\textbf {\bibinfo {volume} {45}},\ \bibinfo {pages} {6479}
  (\bibinfo {year} {1992})}\BibitemShut {NoStop}%
\bibitem [{\citenamefont {Georges}\ \emph {et~al.}(1996)\citenamefont
  {Georges}, \citenamefont {Kotliar}, \citenamefont {Krauth},\ and\
  \citenamefont {Rozenberg}}]{ge.ko.96}%
  \BibitemOpen
  \bibfield  {author} {\bibinfo {author} {\bibfnamefont {A.}~\bibnamefont
  {Georges}}, \bibinfo {author} {\bibfnamefont {G.}~\bibnamefont {Kotliar}},
  \bibinfo {author} {\bibfnamefont {W.}~\bibnamefont {Krauth}},\ and\ \bibinfo
  {author} {\bibfnamefont {M.~J.}\ \bibnamefont {Rozenberg}},\ }\bibfield
  {title} {\bibinfo {title} {Dynamical mean-field theory of strongly correlated
  fermion systems and the limit of infinite dimensions},\ }\href
  {https://doi.org/10.1103/RevModPhys.68.13} {\bibfield  {journal} {\bibinfo
  {journal} {Rev. Mod. Phys.}\ }\textbf {\bibinfo {volume} {68}},\ \bibinfo
  {pages} {13} (\bibinfo {year} {1996})}\BibitemShut {NoStop}%
\bibitem [{\citenamefont {Liebsch}(2003)}]{lieb.03}%
  \BibitemOpen
  \bibfield  {author} {\bibinfo {author} {\bibfnamefont {A.}~\bibnamefont
  {Liebsch}},\ }\bibfield  {title} {\bibinfo {title} {Surface versus bulk
  coulomb correlations in photoemission spectra of
  ${\mathrm{s}\mathrm{r}\mathrm{v}\mathrm{o}}_{3}$ and
  ${\mathrm{c}\mathrm{a}\mathrm{v}\mathrm{o}}_{3}$},\ }\href
  {https://doi.org/10.1103/PhysRevLett.90.096401} {\bibfield  {journal}
  {\bibinfo  {journal} {Phys. Rev. Lett.}\ }\textbf {\bibinfo {volume} {90}},\
  \bibinfo {pages} {096401} (\bibinfo {year} {2003})}\BibitemShut {NoStop}%
\bibitem [{\citenamefont {Sekiyama}\ \emph {et~al.}(2004)\citenamefont
  {Sekiyama}, \citenamefont {Fujiwara}, \citenamefont {Imada}, \citenamefont
  {Suga}, \citenamefont {Eisaki}, \citenamefont {Uchida}, \citenamefont
  {Takegahara}, \citenamefont {Harima}, \citenamefont {Saitoh}, \citenamefont
  {Nekrasov}, \citenamefont {Keller}, \citenamefont {Kondakov}, \citenamefont
  {Kozhevnikov}, \citenamefont {Pruschke}, \citenamefont {Held}, \citenamefont
  {Vollhardt},\ and\ \citenamefont {Anisimov}}]{se.fu.04}%
  \BibitemOpen
  \bibfield  {author} {\bibinfo {author} {\bibfnamefont {A.}~\bibnamefont
  {Sekiyama}}, \bibinfo {author} {\bibfnamefont {H.}~\bibnamefont {Fujiwara}},
  \bibinfo {author} {\bibfnamefont {S.}~\bibnamefont {Imada}}, \bibinfo
  {author} {\bibfnamefont {S.}~\bibnamefont {Suga}}, \bibinfo {author}
  {\bibfnamefont {H.}~\bibnamefont {Eisaki}}, \bibinfo {author} {\bibfnamefont
  {S.~I.}\ \bibnamefont {Uchida}}, \bibinfo {author} {\bibfnamefont
  {K.}~\bibnamefont {Takegahara}}, \bibinfo {author} {\bibfnamefont
  {H.}~\bibnamefont {Harima}}, \bibinfo {author} {\bibfnamefont
  {Y.}~\bibnamefont {Saitoh}}, \bibinfo {author} {\bibfnamefont {I.~A.}\
  \bibnamefont {Nekrasov}}, \bibinfo {author} {\bibfnamefont {G.}~\bibnamefont
  {Keller}}, \bibinfo {author} {\bibfnamefont {D.~E.}\ \bibnamefont
  {Kondakov}}, \bibinfo {author} {\bibfnamefont {A.~V.}\ \bibnamefont
  {Kozhevnikov}}, \bibinfo {author} {\bibfnamefont {T.}~\bibnamefont
  {Pruschke}}, \bibinfo {author} {\bibfnamefont {K.}~\bibnamefont {Held}},
  \bibinfo {author} {\bibfnamefont {D.}~\bibnamefont {Vollhardt}},\ and\
  \bibinfo {author} {\bibfnamefont {V.~I.}\ \bibnamefont {Anisimov}},\
  }\bibfield  {title} {\bibinfo {title} {Mutual experimental and theoretical
  validation of bulk photoemission spectra of
  ${\mathrm{s}\mathrm{r}}_{1\ensuremath{-}x}{\mathrm{c}\mathrm{a}}_{x}{\mathrm{v}\mathrm{o}}_{3}$},\
  }\href {https://doi.org/10.1103/PhysRevLett.93.156402} {\bibfield  {journal}
  {\bibinfo  {journal} {Phys. Rev. Lett.}\ }\textbf {\bibinfo {volume} {93}},\
  \bibinfo {pages} {156402} (\bibinfo {year} {2004})}\BibitemShut {NoStop}%
\bibitem [{\citenamefont {Pavarini}\ \emph {et~al.}(2004)\citenamefont
  {Pavarini}, \citenamefont {Biermann}, \citenamefont {Poteryaev},
  \citenamefont {Lichtenstein}, \citenamefont {Georges},\ and\ \citenamefont
  {Andersen}}]{pa.bi.04.po}%
  \BibitemOpen
  \bibfield  {author} {\bibinfo {author} {\bibfnamefont {E.}~\bibnamefont
  {Pavarini}}, \bibinfo {author} {\bibfnamefont {S.}~\bibnamefont {Biermann}},
  \bibinfo {author} {\bibfnamefont {A.}~\bibnamefont {Poteryaev}}, \bibinfo
  {author} {\bibfnamefont {A.~I.}\ \bibnamefont {Lichtenstein}}, \bibinfo
  {author} {\bibfnamefont {A.}~\bibnamefont {Georges}},\ and\ \bibinfo {author}
  {\bibfnamefont {O.~K.}\ \bibnamefont {Andersen}},\ }\bibfield  {title}
  {\bibinfo {title} {Mott transition and suppression of orbital fluctuations in
  orthorhombic 3d[sup 1] perovskites},\ }\href
  {http://link.aps.org/abstract/PRL/v92/e176403} {\bibfield  {journal}
  {\bibinfo  {journal} {Phys. Rev. Lett.}\ }\textbf {\bibinfo {volume} {92}},\
  \bibinfo {pages} {176403} (\bibinfo {year} {2004})}\BibitemShut {NoStop}%
\bibitem [{\citenamefont {Nekrasov}\ \emph {et~al.}(2005)\citenamefont
  {Nekrasov}, \citenamefont {Keller}, \citenamefont {Kondakov}, \citenamefont
  {Kozhevnikov}, \citenamefont {Pruschke}, \citenamefont {Held}, \citenamefont
  {Vollhardt},\ and\ \citenamefont {Anisimov}}]{ne.ke.05}%
  \BibitemOpen
  \bibfield  {author} {\bibinfo {author} {\bibfnamefont {I.~A.}\ \bibnamefont
  {Nekrasov}}, \bibinfo {author} {\bibfnamefont {G.}~\bibnamefont {Keller}},
  \bibinfo {author} {\bibfnamefont {D.~E.}\ \bibnamefont {Kondakov}}, \bibinfo
  {author} {\bibfnamefont {A.~V.}\ \bibnamefont {Kozhevnikov}}, \bibinfo
  {author} {\bibfnamefont {T.}~\bibnamefont {Pruschke}}, \bibinfo {author}
  {\bibfnamefont {K.}~\bibnamefont {Held}}, \bibinfo {author} {\bibfnamefont
  {D.}~\bibnamefont {Vollhardt}},\ and\ \bibinfo {author} {\bibfnamefont
  {V.~I.}\ \bibnamefont {Anisimov}},\ }\bibfield  {title} {\bibinfo {title}
  {Comparative study of correlation effects in
  $\mathrm{Ca}\mathrm{V}{\mathrm{o}}_{3}$ and
  $\mathrm{Sr}\mathrm{V}{\mathrm{o}}_{3}$},\ }\href
  {https://doi.org/10.1103/PhysRevB.72.155106} {\bibfield  {journal} {\bibinfo
  {journal} {Phys. Rev. B}\ }\textbf {\bibinfo {volume} {72}},\ \bibinfo
  {pages} {155106} (\bibinfo {year} {2005})}\BibitemShut {NoStop}%
\bibitem [{\citenamefont {Nekrasov}\ \emph {et~al.}(2006)\citenamefont
  {Nekrasov}, \citenamefont {Held}, \citenamefont {Keller}, \citenamefont
  {Kondakov}, \citenamefont {Pruschke}, \citenamefont {Kollar}, \citenamefont
  {Andersen}, \citenamefont {Anisimov},\ and\ \citenamefont
  {Vollhardt}}]{ne.he.06}%
  \BibitemOpen
  \bibfield  {author} {\bibinfo {author} {\bibfnamefont {I.~A.}\ \bibnamefont
  {Nekrasov}}, \bibinfo {author} {\bibfnamefont {K.}~\bibnamefont {Held}},
  \bibinfo {author} {\bibfnamefont {G.}~\bibnamefont {Keller}}, \bibinfo
  {author} {\bibfnamefont {D.~E.}\ \bibnamefont {Kondakov}}, \bibinfo {author}
  {\bibfnamefont {T.}~\bibnamefont {Pruschke}}, \bibinfo {author}
  {\bibfnamefont {M.}~\bibnamefont {Kollar}}, \bibinfo {author} {\bibfnamefont
  {O.~K.}\ \bibnamefont {Andersen}}, \bibinfo {author} {\bibfnamefont {V.~I.}\
  \bibnamefont {Anisimov}},\ and\ \bibinfo {author} {\bibfnamefont
  {D.}~\bibnamefont {Vollhardt}},\ }\bibfield  {title} {\bibinfo {title}
  {Momentum-resolved spectral functions of srvo$_{ 3}$ calculated by lda +
  dmft},\ }\href {https://doi.org/10.1103/PhysRevB.73.155112} {\bibfield
  {journal} {\bibinfo  {journal} {Phys. Rev. B}\ }\textbf {\bibinfo {volume}
  {73}},\ \bibinfo {pages} {155112} (\bibinfo {year} {2006})}\BibitemShut
  {NoStop}%
\bibitem [{\citenamefont {Nomura}\ \emph {et~al.}(2012)\citenamefont {Nomura},
  \citenamefont {Kaltak}, \citenamefont {Nakamura}, \citenamefont {Taranto},
  \citenamefont {Sakai}, \citenamefont {Toschi}, \citenamefont {Arita},
  \citenamefont {Held}, \citenamefont {Kresse},\ and\ \citenamefont
  {Imada}}]{no.ka.12}%
  \BibitemOpen
  \bibfield  {author} {\bibinfo {author} {\bibfnamefont {Y.}~\bibnamefont
  {Nomura}}, \bibinfo {author} {\bibfnamefont {M.}~\bibnamefont {Kaltak}},
  \bibinfo {author} {\bibfnamefont {K.}~\bibnamefont {Nakamura}}, \bibinfo
  {author} {\bibfnamefont {C.}~\bibnamefont {Taranto}}, \bibinfo {author}
  {\bibfnamefont {S.}~\bibnamefont {Sakai}}, \bibinfo {author} {\bibfnamefont
  {A.}~\bibnamefont {Toschi}}, \bibinfo {author} {\bibfnamefont
  {R.}~\bibnamefont {Arita}}, \bibinfo {author} {\bibfnamefont
  {K.}~\bibnamefont {Held}}, \bibinfo {author} {\bibfnamefont {G.}~\bibnamefont
  {Kresse}},\ and\ \bibinfo {author} {\bibfnamefont {M.}~\bibnamefont
  {Imada}},\ }\bibfield  {title} {\bibinfo {title} {Effective on-site
  interaction for dynamical mean-field theory},\ }\href
  {https://doi.org/10.1103/PhysRevB.86.085117} {\bibfield  {journal} {\bibinfo
  {journal} {Phys. Rev. B}\ }\textbf {\bibinfo {volume} {86}},\ \bibinfo
  {pages} {085117} (\bibinfo {year} {2012})}\BibitemShut {NoStop}%
\bibitem [{\citenamefont {Inoue}\ \emph {et~al.}(1998)\citenamefont {Inoue},
  \citenamefont {Goto}, \citenamefont {Makino}, \citenamefont {Hussey},\ and\
  \citenamefont {Ishikawa}}]{in.go.98}%
  \BibitemOpen
  \bibfield  {author} {\bibinfo {author} {\bibfnamefont {I.~H.}\ \bibnamefont
  {Inoue}}, \bibinfo {author} {\bibfnamefont {O.}~\bibnamefont {Goto}},
  \bibinfo {author} {\bibfnamefont {H.}~\bibnamefont {Makino}}, \bibinfo
  {author} {\bibfnamefont {N.~E.}\ \bibnamefont {Hussey}},\ and\ \bibinfo
  {author} {\bibfnamefont {M.}~\bibnamefont {Ishikawa}},\ }\bibfield  {title}
  {\bibinfo {title} {Bandwidth control in a perovskite-type
  ${3d}^{1}$-correlated metal
  ${\mathrm{ca}}_{1\ensuremath{-}x}{\mathrm{sr}}_{x}{\mathrm{vo}}_{3}.$ i.
  evolution of the electronic properties and effective mass},\ }\href
  {https://doi.org/10.1103/PhysRevB.58.4372} {\bibfield  {journal} {\bibinfo
  {journal} {Phys. Rev. B}\ }\textbf {\bibinfo {volume} {58}},\ \bibinfo
  {pages} {4372} (\bibinfo {year} {1998})}\BibitemShut {NoStop}%
\bibitem [{\citenamefont {Yoshimatsu}\ \emph {et~al.}(2010)\citenamefont
  {Yoshimatsu}, \citenamefont {Okabe}, \citenamefont {Kumigashira},
  \citenamefont {Okamoto}, \citenamefont {Aizaki}, \citenamefont {Fujimori},\
  and\ \citenamefont {Oshima}}]{yo.ok.10}%
  \BibitemOpen
  \bibfield  {author} {\bibinfo {author} {\bibfnamefont {K.}~\bibnamefont
  {Yoshimatsu}}, \bibinfo {author} {\bibfnamefont {T.}~\bibnamefont {Okabe}},
  \bibinfo {author} {\bibfnamefont {H.}~\bibnamefont {Kumigashira}}, \bibinfo
  {author} {\bibfnamefont {S.}~\bibnamefont {Okamoto}}, \bibinfo {author}
  {\bibfnamefont {S.}~\bibnamefont {Aizaki}}, \bibinfo {author} {\bibfnamefont
  {A.}~\bibnamefont {Fujimori}},\ and\ \bibinfo {author} {\bibfnamefont
  {M.}~\bibnamefont {Oshima}},\ }\bibfield  {title} {\bibinfo {title}
  {Dimensional-crossover-driven metal-insulator transition in
  ${\mathrm{srvo}}_{3}$ ultrathin films},\ }\href
  {https://doi.org/10.1103/PhysRevLett.104.147601} {\bibfield  {journal}
  {\bibinfo  {journal} {Phys. Rev. Lett.}\ }\textbf {\bibinfo {volume} {104}},\
  \bibinfo {pages} {147601} (\bibinfo {year} {2010})}\BibitemShut {NoStop}%
\bibitem [{\citenamefont {Zhong}\ \emph {et~al.}(2015)\citenamefont {Zhong},
  \citenamefont {Wallerberger}, \citenamefont {Tomczak}, \citenamefont
  {Taranto}, \citenamefont {Parragh}, \citenamefont {Toschi}, \citenamefont
  {Sangiovanni},\ and\ \citenamefont {Held}}]{zh.wa.15}%
  \BibitemOpen
  \bibfield  {author} {\bibinfo {author} {\bibfnamefont {Z.}~\bibnamefont
  {Zhong}}, \bibinfo {author} {\bibfnamefont {M.}~\bibnamefont {Wallerberger}},
  \bibinfo {author} {\bibfnamefont {J.~M.}\ \bibnamefont {Tomczak}}, \bibinfo
  {author} {\bibfnamefont {C.}~\bibnamefont {Taranto}}, \bibinfo {author}
  {\bibfnamefont {N.}~\bibnamefont {Parragh}}, \bibinfo {author} {\bibfnamefont
  {A.}~\bibnamefont {Toschi}}, \bibinfo {author} {\bibfnamefont
  {G.}~\bibnamefont {Sangiovanni}},\ and\ \bibinfo {author} {\bibfnamefont
  {K.}~\bibnamefont {Held}},\ }\bibfield  {title} {\bibinfo {title}
  {Electronics with correlated oxides:
  ${\mathrm{srvo}}_{3}/{\mathrm{srtio}}_{3}$ as a mott transistor},\ }\href
  {https://doi.org/10.1103/PhysRevLett.114.246401} {\bibfield  {journal}
  {\bibinfo  {journal} {Phys. Rev. Lett.}\ }\textbf {\bibinfo {volume} {114}},\
  \bibinfo {pages} {246401} (\bibinfo {year} {2015})}\BibitemShut {NoStop}%
\bibitem [{\citenamefont {Bhandary}\ \emph {et~al.}(2016)\citenamefont
  {Bhandary}, \citenamefont {Assmann}, \citenamefont {Aichhorn},\ and\
  \citenamefont {Held}}]{ba.as.16}%
  \BibitemOpen
  \bibfield  {author} {\bibinfo {author} {\bibfnamefont {S.}~\bibnamefont
  {Bhandary}}, \bibinfo {author} {\bibfnamefont {E.}~\bibnamefont {Assmann}},
  \bibinfo {author} {\bibfnamefont {M.}~\bibnamefont {Aichhorn}},\ and\
  \bibinfo {author} {\bibfnamefont {K.}~\bibnamefont {Held}},\ }\bibfield
  {title} {\bibinfo {title} {Charge self-consistency in density functional
  theory combined with dynamical mean field theory: $k$-space reoccupation and
  orbital order},\ }\href {https://doi.org/10.1103/PhysRevB.94.155131}
  {\bibfield  {journal} {\bibinfo  {journal} {Phys. Rev. B}\ }\textbf {\bibinfo
  {volume} {94}},\ \bibinfo {pages} {155131} (\bibinfo {year}
  {2016})}\BibitemShut {NoStop}%
\bibitem [{\citenamefont {Murakami}\ and\ \citenamefont
  {Werner}(2018)}]{mu.we.18}%
  \BibitemOpen
  \bibfield  {author} {\bibinfo {author} {\bibfnamefont {Y.}~\bibnamefont
  {Murakami}}\ and\ \bibinfo {author} {\bibfnamefont {P.}~\bibnamefont
  {Werner}},\ }\bibfield  {title} {\bibinfo {title} {Nonequilibrium steady
  states of electric field driven mott insulators},\ }\href
  {https://doi.org/10.1103/PhysRevB.98.075102} {\bibfield  {journal} {\bibinfo
  {journal} {Phys. Rev. B}\ }\textbf {\bibinfo {volume} {98}},\ \bibinfo
  {pages} {075102} (\bibinfo {year} {2018})}\BibitemShut {NoStop}%
\bibitem [{\citenamefont {Picano}\ \emph {et~al.}(2021)\citenamefont {Picano},
  \citenamefont {Li},\ and\ \citenamefont {Eckstein}}]{pi.li.21}%
  \BibitemOpen
  \bibfield  {author} {\bibinfo {author} {\bibfnamefont {A.}~\bibnamefont
  {Picano}}, \bibinfo {author} {\bibfnamefont {J.}~\bibnamefont {Li}},\ and\
  \bibinfo {author} {\bibfnamefont {M.}~\bibnamefont {Eckstein}},\ }\bibfield
  {title} {\bibinfo {title} {Quantum boltzmann equation for strongly correlated
  electrons},\ }\href@noop {} {\bibfield  {journal} {\bibinfo  {journal} {Phys.
  Rev. B}\ }\textbf {\bibinfo {volume} {104}},\ \bibinfo {pages} {085108}
  (\bibinfo {year} {2021})}\BibitemShut {NoStop}%
\bibitem [{\citenamefont {Mazzocchi}\ \emph {et~al.}(2022)\citenamefont
  {Mazzocchi}, \citenamefont {Gazzaneo}, \citenamefont {Lotze},\ and\
  \citenamefont {Arrigoni}}]{ma.ga.22}%
  \BibitemOpen
  \bibfield  {author} {\bibinfo {author} {\bibfnamefont {T.~M.}\ \bibnamefont
  {Mazzocchi}}, \bibinfo {author} {\bibfnamefont {P.}~\bibnamefont {Gazzaneo}},
  \bibinfo {author} {\bibfnamefont {J.}~\bibnamefont {Lotze}},\ and\ \bibinfo
  {author} {\bibfnamefont {E.}~\bibnamefont {Arrigoni}},\ }\bibfield  {title}
  {\bibinfo {title} {Correlated mott insulators in strong electric fields: Role
  of phonons in heat dissipation},\ }\href
  {https://doi.org/10.1103/PhysRevB.106.125123} {\bibfield  {journal} {\bibinfo
   {journal} {Phys. Rev. B}\ }\textbf {\bibinfo {volume} {106}},\ \bibinfo
  {pages} {125123} (\bibinfo {year} {2022})}\BibitemShut {NoStop}%
\bibitem [{\citenamefont {Gazzaneo}\ \emph {et~al.}(2022)\citenamefont
  {Gazzaneo}, \citenamefont {Mazzocchi}, \citenamefont {Lotze},\ and\
  \citenamefont {Arrigoni}}]{ga.ma.22}%
  \BibitemOpen
  \bibfield  {author} {\bibinfo {author} {\bibfnamefont {P.}~\bibnamefont
  {Gazzaneo}}, \bibinfo {author} {\bibfnamefont {T.~M.}\ \bibnamefont
  {Mazzocchi}}, \bibinfo {author} {\bibfnamefont {J.}~\bibnamefont {Lotze}},\
  and\ \bibinfo {author} {\bibfnamefont {E.}~\bibnamefont {Arrigoni}},\
  }\bibfield  {title} {\bibinfo {title} {Impact ionization processes in a
  photodriven mott insulator: Influence of phononic dissipation},\ }\href
  {https://doi.org/10.1103/PhysRevB.106.195140} {\bibfield  {journal} {\bibinfo
   {journal} {Phys. Rev. B}\ }\textbf {\bibinfo {volume} {106}},\ \bibinfo
  {pages} {195140} (\bibinfo {year} {2022})}\BibitemShut {NoStop}%
\bibitem [{\citenamefont {Han}\ \emph {et~al.}(2023)\citenamefont {Han},
  \citenamefont {Aron}, \citenamefont {Chen}, \citenamefont {Mansaray},
  \citenamefont {Han}, \citenamefont {Kim}, \citenamefont {Randle},\ and\
  \citenamefont {Bird}}]{ha.ar.23}%
  \BibitemOpen
  \bibfield  {author} {\bibinfo {author} {\bibfnamefont {J.~E.}\ \bibnamefont
  {Han}}, \bibinfo {author} {\bibfnamefont {C.}~\bibnamefont {Aron}}, \bibinfo
  {author} {\bibfnamefont {X.}~\bibnamefont {Chen}}, \bibinfo {author}
  {\bibfnamefont {I.}~\bibnamefont {Mansaray}}, \bibinfo {author}
  {\bibfnamefont {J.-H.}\ \bibnamefont {Han}}, \bibinfo {author} {\bibfnamefont
  {K.-S.}\ \bibnamefont {Kim}}, \bibinfo {author} {\bibfnamefont
  {M.}~\bibnamefont {Randle}},\ and\ \bibinfo {author} {\bibfnamefont {J.~P.}\
  \bibnamefont {Bird}},\ }\bibfield  {title} {\bibinfo {title} {Correlated
  insulator collapse due to quantum avalanche via in-gap ladder states},\
  }\bibfield  {journal} {\bibinfo  {journal} {Nature Communications}\ }\textbf
  {\bibinfo {volume} {14}},\ \href {https://doi.org/10.1038/s41467-023-38557-8}
  {10.1038/s41467-023-38557-8} (\bibinfo {year} {2023})\BibitemShut {NoStop}%
\bibitem [{\citenamefont {Gazzaneo}\ \emph {et~al.}(2024)\citenamefont
  {Gazzaneo}, \citenamefont {Werner}, \citenamefont {Mazzocchi},\ and\
  \citenamefont {Arrigoni}}]{ga.we.24}%
  \BibitemOpen
  \bibfield  {author} {\bibinfo {author} {\bibfnamefont {P.}~\bibnamefont
  {Gazzaneo}}, \bibinfo {author} {\bibfnamefont {D.}~\bibnamefont {Werner}},
  \bibinfo {author} {\bibfnamefont {T.~M.}\ \bibnamefont {Mazzocchi}},\ and\
  \bibinfo {author} {\bibfnamefont {E.}~\bibnamefont {Arrigoni}},\ }\bibfield
  {title} {\bibinfo {title} {Photodriven mott insulating heterostructures: A
  steady-state study of impact ionization processes},\ }\href
  {https://doi.org/10.1103/PhysRevB.109.235134} {\bibfield  {journal} {\bibinfo
   {journal} {Phys. Rev. B}\ }\textbf {\bibinfo {volume} {109}},\ \bibinfo
  {pages} {235134} (\bibinfo {year} {2024})}\BibitemShut {NoStop}%
\bibitem [{\citenamefont {Dohner}\ \emph {et~al.}(2022)\citenamefont {Dohner},
  \citenamefont {Terletska}, \citenamefont {Tam}, \citenamefont {Moreno},\ and\
  \citenamefont {Fotso}}]{do.te.22}%
  \BibitemOpen
  \bibfield  {author} {\bibinfo {author} {\bibfnamefont {E.}~\bibnamefont
  {Dohner}}, \bibinfo {author} {\bibfnamefont {H.}~\bibnamefont {Terletska}},
  \bibinfo {author} {\bibfnamefont {K.-M.}\ \bibnamefont {Tam}}, \bibinfo
  {author} {\bibfnamefont {J.}~\bibnamefont {Moreno}},\ and\ \bibinfo {author}
  {\bibfnamefont {H.~F.}\ \bibnamefont {Fotso}},\ }\bibfield  {title} {\bibinfo
  {title} {Nonequilibrium $\text{DMFT}+\text{CPA}$ for correlated disordered
  systems},\ }\href {https://doi.org/10.1103/PhysRevB.106.195156} {\bibfield
  {journal} {\bibinfo  {journal} {Phys. Rev. B}\ }\textbf {\bibinfo {volume}
  {106}},\ \bibinfo {pages} {195156} (\bibinfo {year} {2022})}\BibitemShut
  {NoStop}%
\bibitem [{\citenamefont {Mazzocchi}\ \emph {et~al.}(2023)\citenamefont
  {Mazzocchi}, \citenamefont {Werner}, \citenamefont {Gazzaneo},\ and\
  \citenamefont {Arrigoni}}]{ma.we.23}%
  \BibitemOpen
  \bibfield  {author} {\bibinfo {author} {\bibfnamefont {T.~M.}\ \bibnamefont
  {Mazzocchi}}, \bibinfo {author} {\bibfnamefont {D.}~\bibnamefont {Werner}},
  \bibinfo {author} {\bibfnamefont {P.}~\bibnamefont {Gazzaneo}},\ and\
  \bibinfo {author} {\bibfnamefont {E.}~\bibnamefont {Arrigoni}},\ }\bibfield
  {title} {\bibinfo {title} {Correlated mott insulators in a strong electric
  field: The effects of phonon renormalization},\ }\href
  {https://doi.org/10.1103/PhysRevB.107.155103} {\bibfield  {journal} {\bibinfo
   {journal} {Phys. Rev. B}\ }\textbf {\bibinfo {volume} {107}},\ \bibinfo
  {pages} {155103} (\bibinfo {year} {2023})}\BibitemShut {NoStop}%
\bibitem [{\citenamefont {Aoki}\ \emph {et~al.}(2014)\citenamefont {Aoki},
  \citenamefont {Tsuji}, \citenamefont {Eckstein}, \citenamefont {Kollar},
  \citenamefont {Oka},\ and\ \citenamefont {Werner}}]{ao.ts.14}%
  \BibitemOpen
  \bibfield  {author} {\bibinfo {author} {\bibfnamefont {H.}~\bibnamefont
  {Aoki}}, \bibinfo {author} {\bibfnamefont {N.}~\bibnamefont {Tsuji}},
  \bibinfo {author} {\bibfnamefont {M.}~\bibnamefont {Eckstein}}, \bibinfo
  {author} {\bibfnamefont {M.}~\bibnamefont {Kollar}}, \bibinfo {author}
  {\bibfnamefont {T.}~\bibnamefont {Oka}},\ and\ \bibinfo {author}
  {\bibfnamefont {P.}~\bibnamefont {Werner}},\ }\bibfield  {title} {\bibinfo
  {title} {Nonequilibrium dynamical mean-field theory and its applications},\
  }\href {https://doi.org/10.1103/RevModPhys.86.779} {\bibfield  {journal}
  {\bibinfo  {journal} {Rev. Mod. Phys.}\ }\textbf {\bibinfo {volume} {86}},\
  \bibinfo {pages} {779} (\bibinfo {year} {2014})}\BibitemShut {NoStop}%
\bibitem [{\citenamefont {Gubernatis}\ \emph {et~al.}(1991)\citenamefont
  {Gubernatis}, \citenamefont {Jarrell}, \citenamefont {Silver},\ and\
  \citenamefont {Sivia}}]{gu.ja.91}%
  \BibitemOpen
  \bibfield  {author} {\bibinfo {author} {\bibfnamefont {J.~E.}\ \bibnamefont
  {Gubernatis}}, \bibinfo {author} {\bibfnamefont {M.}~\bibnamefont {Jarrell}},
  \bibinfo {author} {\bibfnamefont {R.~N.}\ \bibnamefont {Silver}},\ and\
  \bibinfo {author} {\bibfnamefont {D.~S.}\ \bibnamefont {Sivia}},\ }\bibfield
  {title} {\bibinfo {title} {Quantum monte carlo simulations and maximum
  entropy: Dynamics from imaginary-time data},\ }\href
  {https://doi.org/10.1103/PhysRevB.44.6011} {\bibfield  {journal} {\bibinfo
  {journal} {Phys. Rev. B}\ }\textbf {\bibinfo {volume} {44}},\ \bibinfo
  {pages} {6011} (\bibinfo {year} {1991})}\BibitemShut {NoStop}%
\bibitem [{\citenamefont {Wilson}(1975)}]{wils.75}%
  \BibitemOpen
  \bibfield  {author} {\bibinfo {author} {\bibfnamefont {K.~G.}\ \bibnamefont
  {Wilson}},\ }\bibfield  {title} {\bibinfo {title} {The renormalization group:
  Critical phenomena and the kondo problem},\ }\href@noop {} {\bibfield
  {journal} {\bibinfo  {journal} {Rev. Mod. Phys.}\ }\textbf {\bibinfo {volume}
  {47}},\ \bibinfo {pages} {773} (\bibinfo {year} {1975})}\BibitemShut
  {NoStop}%
\bibitem [{\citenamefont {Dorda}\ \emph {et~al.}(2014)\citenamefont {Dorda},
  \citenamefont {Nuss}, \citenamefont {von~der Linden},\ and\ \citenamefont
  {Arrigoni}}]{do.nu.14}%
  \BibitemOpen
  \bibfield  {author} {\bibinfo {author} {\bibfnamefont {A.}~\bibnamefont
  {Dorda}}, \bibinfo {author} {\bibfnamefont {M.}~\bibnamefont {Nuss}},
  \bibinfo {author} {\bibfnamefont {W.}~\bibnamefont {von~der Linden}},\ and\
  \bibinfo {author} {\bibfnamefont {E.}~\bibnamefont {Arrigoni}},\ }\bibfield
  {title} {\bibinfo {title} {Auxiliary master equation approach to non --
  equilibrium correlated impurities},\ }\href
  {https://doi.org/10.1103/PhysRevB.89.165105} {\bibfield  {journal} {\bibinfo
  {journal} {Phys. Rev. B}\ }\textbf {\bibinfo {volume} {89}},\ \bibinfo
  {pages} {165105} (\bibinfo {year} {2014})}\BibitemShut {NoStop}%
\bibitem [{\citenamefont {Werner}\ \emph {et~al.}(2023)\citenamefont {Werner},
  \citenamefont {Lotze},\ and\ \citenamefont {Arrigoni}}]{we.lo.23}%
  \BibitemOpen
  \bibfield  {author} {\bibinfo {author} {\bibfnamefont {D.}~\bibnamefont
  {Werner}}, \bibinfo {author} {\bibfnamefont {J.}~\bibnamefont {Lotze}},\ and\
  \bibinfo {author} {\bibfnamefont {E.}~\bibnamefont {Arrigoni}},\ }\bibfield
  {title} {\bibinfo {title} {Configuration interaction based nonequilibrium
  steady state impurity solver},\ }\href
  {https://doi.org/10.1103/PhysRevB.107.075119} {\bibfield  {journal} {\bibinfo
   {journal} {Phys. Rev. B}\ }\textbf {\bibinfo {volume} {107}},\ \bibinfo
  {pages} {075119} (\bibinfo {year} {2023})}\BibitemShut {NoStop}%
\bibitem [{\citenamefont {Gull}\ \emph {et~al.}(2011)\citenamefont {Gull},
  \citenamefont {Millis}, \citenamefont {Lichtenstein}, \citenamefont
  {Rubtsov}, \citenamefont {Troyer},\ and\ \citenamefont {Werner}}]{gu.mi.11}%
  \BibitemOpen
  \bibfield  {author} {\bibinfo {author} {\bibfnamefont {E.}~\bibnamefont
  {Gull}}, \bibinfo {author} {\bibfnamefont {A.~J.}\ \bibnamefont {Millis}},
  \bibinfo {author} {\bibfnamefont {A.~I.}\ \bibnamefont {Lichtenstein}},
  \bibinfo {author} {\bibfnamefont {A.~N.}\ \bibnamefont {Rubtsov}}, \bibinfo
  {author} {\bibfnamefont {M.}~\bibnamefont {Troyer}},\ and\ \bibinfo {author}
  {\bibfnamefont {P.}~\bibnamefont {Werner}},\ }\bibfield  {title} {\bibinfo
  {title} {Continuous -- time monte carlo methods for quantum impurity
  models},\ }\href@noop {} {\bibfield  {journal} {\bibinfo  {journal} {Rev.
  Mod. Phys.}\ }\textbf {\bibinfo {volume} {83}},\ \bibinfo {pages} {349}
  (\bibinfo {year} {2011})}\BibitemShut {NoStop}%
\bibitem [{\citenamefont {Liebsch}\ and\ \citenamefont
  {Ishida}(2011)}]{li.is.11}%
  \BibitemOpen
  \bibfield  {author} {\bibinfo {author} {\bibfnamefont {A.}~\bibnamefont
  {Liebsch}}\ and\ \bibinfo {author} {\bibfnamefont {H.}~\bibnamefont
  {Ishida}},\ }\bibfield  {title} {\bibinfo {title} {Temperature and bath size
  in exact diagonalization dynamical mean field theory},\ }\href
  {https://doi.org/10.1088/0953-8984/24/5/053201} {\bibfield  {journal}
  {\bibinfo  {journal} {Journal of Physics: Condensed Matter}\ }\textbf
  {\bibinfo {volume} {24}},\ \bibinfo {pages} {053201} (\bibinfo {year}
  {2011})}\BibitemShut {NoStop}%
\bibitem [{\citenamefont {Crippa}\ \emph {et~al.}(2025)\citenamefont {Crippa},
  \citenamefont {Krivenko}, \citenamefont {Giuli}, \citenamefont {Bellomia},
  \citenamefont {Kowalski}, \citenamefont {Petocchi}, \citenamefont {Scazzola},
  \citenamefont {Wallerberger}, \citenamefont {Mazza}, \citenamefont
  {de~Medici}, \citenamefont {Sangiovanni}, \citenamefont {Capone},\ and\
  \citenamefont {Amaricci}}]{cr.kr.25}%
  \BibitemOpen
  \bibfield  {author} {\bibinfo {author} {\bibfnamefont {L.}~\bibnamefont
  {Crippa}}, \bibinfo {author} {\bibfnamefont {I.}~\bibnamefont {Krivenko}},
  \bibinfo {author} {\bibfnamefont {S.}~\bibnamefont {Giuli}}, \bibinfo
  {author} {\bibfnamefont {G.}~\bibnamefont {Bellomia}}, \bibinfo {author}
  {\bibfnamefont {A.}~\bibnamefont {Kowalski}}, \bibinfo {author}
  {\bibfnamefont {F.}~\bibnamefont {Petocchi}}, \bibinfo {author}
  {\bibfnamefont {A.}~\bibnamefont {Scazzola}}, \bibinfo {author}
  {\bibfnamefont {M.}~\bibnamefont {Wallerberger}}, \bibinfo {author}
  {\bibfnamefont {G.}~\bibnamefont {Mazza}}, \bibinfo {author} {\bibfnamefont
  {L.}~\bibnamefont {de~Medici}}, \bibinfo {author} {\bibfnamefont
  {G.}~\bibnamefont {Sangiovanni}}, \bibinfo {author} {\bibfnamefont
  {M.}~\bibnamefont {Capone}},\ and\ \bibinfo {author} {\bibfnamefont
  {A.}~\bibnamefont {Amaricci}},\ }\href {https://arxiv.org/abs/2506.01363} {}
  (\bibinfo {year} {2025}),\ \Eprint {https://arxiv.org/abs/2506.01363}
  {arXiv:2506.01363 [cond-mat.str-el]} \BibitemShut {NoStop}%
\bibitem [{\citenamefont {Mitchell}\ \emph {et~al.}(2014)\citenamefont
  {Mitchell}, \citenamefont {Galpin}, \citenamefont {Wilson-Fletcher},
  \citenamefont {Logan},\ and\ \citenamefont {Bulla}}]{mi.ga.14}%
  \BibitemOpen
  \bibfield  {author} {\bibinfo {author} {\bibfnamefont {A.~K.}\ \bibnamefont
  {Mitchell}}, \bibinfo {author} {\bibfnamefont {M.~R.}\ \bibnamefont
  {Galpin}}, \bibinfo {author} {\bibfnamefont {S.}~\bibnamefont
  {Wilson-Fletcher}}, \bibinfo {author} {\bibfnamefont {D.~E.}\ \bibnamefont
  {Logan}},\ and\ \bibinfo {author} {\bibfnamefont {R.}~\bibnamefont {Bulla}},\
  }\bibfield  {title} {\bibinfo {title} {Generalized wilson chain for solving
  multichannel quantum impurity problems},\ }\href
  {https://doi.org/10.1103/PhysRevB.89.121105} {\bibfield  {journal} {\bibinfo
  {journal} {Phys. Rev. B}\ }\textbf {\bibinfo {volume} {89}},\ \bibinfo
  {pages} {121105} (\bibinfo {year} {2014})}\BibitemShut {NoStop}%
\bibitem [{\citenamefont {Stadler}\ \emph {et~al.}(2016)\citenamefont
  {Stadler}, \citenamefont {Mitchell}, \citenamefont {von Delft},\ and\
  \citenamefont {Weichselbaum}}]{st.mi.16}%
  \BibitemOpen
  \bibfield  {author} {\bibinfo {author} {\bibfnamefont {K.~M.}\ \bibnamefont
  {Stadler}}, \bibinfo {author} {\bibfnamefont {A.~K.}\ \bibnamefont
  {Mitchell}}, \bibinfo {author} {\bibfnamefont {J.}~\bibnamefont {von
  Delft}},\ and\ \bibinfo {author} {\bibfnamefont {A.}~\bibnamefont
  {Weichselbaum}},\ }\bibfield  {title} {\bibinfo {title} {Interleaved
  numerical renormalization group as an efficient multiband impurity solver},\
  }\href {https://doi.org/10.1103/PhysRevB.93.235101} {\bibfield  {journal}
  {\bibinfo  {journal} {Phys. Rev. B}\ }\textbf {\bibinfo {volume} {93}},\
  \bibinfo {pages} {235101} (\bibinfo {year} {2016})}\BibitemShut {NoStop}%
\bibitem [{\citenamefont {Werner}\ \emph {et~al.}(2006)\citenamefont {Werner},
  \citenamefont {Comanac}, \citenamefont {de' Medici}, \citenamefont {Troyer},\
  and\ \citenamefont {Millis}}]{we.co.06}%
  \BibitemOpen
  \bibfield  {author} {\bibinfo {author} {\bibfnamefont {P.}~\bibnamefont
  {Werner}}, \bibinfo {author} {\bibfnamefont {A.}~\bibnamefont {Comanac}},
  \bibinfo {author} {\bibfnamefont {L.}~\bibnamefont {de' Medici}}, \bibinfo
  {author} {\bibfnamefont {M.}~\bibnamefont {Troyer}},\ and\ \bibinfo {author}
  {\bibfnamefont {A.~J.}\ \bibnamefont {Millis}},\ }\bibfield  {title}
  {\bibinfo {title} {Continuous-time solver for quantum impurity models},\
  }\href {https://doi.org/10.1103/PhysRevLett.97.076405} {\bibfield  {journal}
  {\bibinfo  {journal} {Phys. Rev. Lett.}\ }\textbf {\bibinfo {volume} {97}},\
  \bibinfo {pages} {076405} (\bibinfo {year} {2006})}\BibitemShut {NoStop}%
\bibitem [{\citenamefont {Aichhorn}\ \emph {et~al.}(2010)\citenamefont
  {Aichhorn}, \citenamefont {Biermann}, \citenamefont {Miyake}, \citenamefont
  {Georges},\ and\ \citenamefont {Imada}}]{ai.bi.10}%
  \BibitemOpen
  \bibfield  {author} {\bibinfo {author} {\bibfnamefont {M.}~\bibnamefont
  {Aichhorn}}, \bibinfo {author} {\bibfnamefont {S.}~\bibnamefont {Biermann}},
  \bibinfo {author} {\bibfnamefont {T.}~\bibnamefont {Miyake}}, \bibinfo
  {author} {\bibfnamefont {A.}~\bibnamefont {Georges}},\ and\ \bibinfo {author}
  {\bibfnamefont {M.}~\bibnamefont {Imada}},\ }\bibfield  {title} {\bibinfo
  {title} {Theoretical evidence for strong correlations and incoherent metallic
  state in fese},\ }\href {https://doi.org/10.1103/PhysRevB.82.064504}
  {\bibfield  {journal} {\bibinfo  {journal} {Phys. Rev. B}\ }\textbf {\bibinfo
  {volume} {82}},\ \bibinfo {pages} {064504} (\bibinfo {year}
  {2010})}\BibitemShut {NoStop}%
\bibitem [{\citenamefont {Yin}\ \emph {et~al.}(2011)\citenamefont {Yin},
  \citenamefont {Haule},\ and\ \citenamefont {Kotliar}}]{yi.ha.11}%
  \BibitemOpen
  \bibfield  {author} {\bibinfo {author} {\bibfnamefont {Z.~P.}\ \bibnamefont
  {Yin}}, \bibinfo {author} {\bibfnamefont {K.}~\bibnamefont {Haule}},\ and\
  \bibinfo {author} {\bibfnamefont {G.}~\bibnamefont {Kotliar}},\ }\bibfield
  {title} {\bibinfo {title} {Kinetic frustration and the nature of the magnetic
  and paramagnetic states in iron pnictides and iron chalcogenides},\
  }\bibfield  {journal} {\bibinfo  {journal} {Nature Materials}\ }\textbf
  {\bibinfo {volume} {10}},\ \href {https://doi.org/10.1038/nmat3120}
  {10.1038/nmat3120} (\bibinfo {year} {2011})\BibitemShut {NoStop}%
\bibitem [{\citenamefont {Eidelstein}\ \emph {et~al.}(2020)\citenamefont
  {Eidelstein}, \citenamefont {Gull},\ and\ \citenamefont {Cohen}}]{ei.gu.20}%
  \BibitemOpen
  \bibfield  {author} {\bibinfo {author} {\bibfnamefont {E.}~\bibnamefont
  {Eidelstein}}, \bibinfo {author} {\bibfnamefont {E.}~\bibnamefont {Gull}},\
  and\ \bibinfo {author} {\bibfnamefont {G.}~\bibnamefont {Cohen}},\ }\bibfield
   {title} {\bibinfo {title} {Multiorbital quantum impurity solver for general
  interactions and hybridizations},\ }\href
  {https://doi.org/10.1103/PhysRevLett.124.206405} {\bibfield  {journal}
  {\bibinfo  {journal} {Phys. Rev. Lett.}\ }\textbf {\bibinfo {volume} {124}},\
  \bibinfo {pages} {206405} (\bibinfo {year} {2020})}\BibitemShut {NoStop}%
\bibitem [{\citenamefont {N\'u\~nez Fern\'andez}\ \emph
  {et~al.}(2022)\citenamefont {N\'u\~nez Fern\'andez}, \citenamefont {Jeannin},
  \citenamefont {Dumitrescu}, \citenamefont {Kloss}, \citenamefont {Kaye},
  \citenamefont {Parcollet},\ and\ \citenamefont {Waintal}}]{nu.je.22}%
  \BibitemOpen
  \bibfield  {author} {\bibinfo {author} {\bibfnamefont {Y.}~\bibnamefont
  {N\'u\~nez Fern\'andez}}, \bibinfo {author} {\bibfnamefont {M.}~\bibnamefont
  {Jeannin}}, \bibinfo {author} {\bibfnamefont {P.~T.}\ \bibnamefont
  {Dumitrescu}}, \bibinfo {author} {\bibfnamefont {T.}~\bibnamefont {Kloss}},
  \bibinfo {author} {\bibfnamefont {J.}~\bibnamefont {Kaye}}, \bibinfo {author}
  {\bibfnamefont {O.}~\bibnamefont {Parcollet}},\ and\ \bibinfo {author}
  {\bibfnamefont {X.}~\bibnamefont {Waintal}},\ }\bibfield  {title} {\bibinfo
  {title} {Learning feynman diagrams with tensor trains},\ }\href
  {https://doi.org/10.1103/PhysRevX.12.041018} {\bibfield  {journal} {\bibinfo
  {journal} {Phys. Rev. X}\ }\textbf {\bibinfo {volume} {12}},\ \bibinfo
  {pages} {041018} (\bibinfo {year} {2022})}\BibitemShut {NoStop}%
\bibitem [{\citenamefont {Cohen}\ \emph {et~al.}(2015)\citenamefont {Cohen},
  \citenamefont {Gull}, \citenamefont {Reichman},\ and\ \citenamefont
  {Millis}}]{co.gu.15}%
  \BibitemOpen
  \bibfield  {author} {\bibinfo {author} {\bibfnamefont {G.}~\bibnamefont
  {Cohen}}, \bibinfo {author} {\bibfnamefont {E.}~\bibnamefont {Gull}},
  \bibinfo {author} {\bibfnamefont {D.~R.}\ \bibnamefont {Reichman}},\ and\
  \bibinfo {author} {\bibfnamefont {A.~J.}\ \bibnamefont {Millis}},\ }\bibfield
   {title} {\bibinfo {title} {Taming the dynamical sign problem in real-time
  evolution of quantum many-body problems},\ }\href
  {https://doi.org/10.1103/PhysRevLett.115.266802} {\bibfield  {journal}
  {\bibinfo  {journal} {Phys. Rev. Lett.}\ }\textbf {\bibinfo {volume} {115}},\
  \bibinfo {pages} {266802} (\bibinfo {year} {2015})}\BibitemShut {NoStop}%
\bibitem [{\citenamefont {Erpenbeck}\ \emph {et~al.}(2023)\citenamefont
  {Erpenbeck}, \citenamefont {Gull},\ and\ \citenamefont {Cohen}}]{er.gu.23}%
  \BibitemOpen
  \bibfield  {author} {\bibinfo {author} {\bibfnamefont {A.}~\bibnamefont
  {Erpenbeck}}, \bibinfo {author} {\bibfnamefont {E.}~\bibnamefont {Gull}},\
  and\ \bibinfo {author} {\bibfnamefont {G.}~\bibnamefont {Cohen}},\ }\bibfield
   {title} {\bibinfo {title} {Quantum monte carlo method in the steady state},\
  }\href {https://doi.org/10.1103/PhysRevLett.130.186301} {\bibfield  {journal}
  {\bibinfo  {journal} {Phys. Rev. Lett.}\ }\textbf {\bibinfo {volume} {130}},\
  \bibinfo {pages} {186301} (\bibinfo {year} {2023})}\BibitemShut {NoStop}%
\bibitem [{\citenamefont {Erpenbeck}\ \emph {et~al.}(2024)\citenamefont
  {Erpenbeck}, \citenamefont {Blommel}, \citenamefont {Zhang}, \citenamefont
  {Lin}, \citenamefont {Cohen},\ and\ \citenamefont {Gull}}]{er.bl.24}%
  \BibitemOpen
  \bibfield  {author} {\bibinfo {author} {\bibfnamefont {A.}~\bibnamefont
  {Erpenbeck}}, \bibinfo {author} {\bibfnamefont {T.}~\bibnamefont {Blommel}},
  \bibinfo {author} {\bibfnamefont {L.}~\bibnamefont {Zhang}}, \bibinfo
  {author} {\bibfnamefont {W.-T.}\ \bibnamefont {Lin}}, \bibinfo {author}
  {\bibfnamefont {G.}~\bibnamefont {Cohen}},\ and\ \bibinfo {author}
  {\bibfnamefont {E.}~\bibnamefont {Gull}},\ }\bibfield  {title} {\bibinfo
  {title} {Steady-state properties of multi-orbital systems using quantum monte
  carlo},\ }\href {https://doi.org/10.1063/5.0226253} {\bibfield  {journal}
  {\bibinfo  {journal} {The Journal of Chemical Physics}\ }\textbf {\bibinfo
  {volume} {161}},\ \bibinfo {pages} {094104} (\bibinfo {year} {2024})},\
  \Eprint
  {https://arxiv.org/abs/https://pubs.aip.org/aip/jcp/article-pdf/doi/10.1063/5.0226253/20142171/094104\_1\_5.0226253.pdf}
  {https://pubs.aip.org/aip/jcp/article-pdf/doi/10.1063/5.0226253/20142171/094104\_1\_5.0226253.pdf}
  \BibitemShut {NoStop}%
\bibitem [{\citenamefont {K\"unzel}\ \emph {et~al.}(2024)\citenamefont
  {K\"unzel}, \citenamefont {Erpenbeck}, \citenamefont {Werner}, \citenamefont
  {Arrigoni}, \citenamefont {Gull}, \citenamefont {Cohen},\ and\ \citenamefont
  {Eckstein}}]{ku.er.24}%
  \BibitemOpen
  \bibfield  {author} {\bibinfo {author} {\bibfnamefont {F.}~\bibnamefont
  {K\"unzel}}, \bibinfo {author} {\bibfnamefont {A.}~\bibnamefont {Erpenbeck}},
  \bibinfo {author} {\bibfnamefont {D.}~\bibnamefont {Werner}}, \bibinfo
  {author} {\bibfnamefont {E.}~\bibnamefont {Arrigoni}}, \bibinfo {author}
  {\bibfnamefont {E.}~\bibnamefont {Gull}}, \bibinfo {author} {\bibfnamefont
  {G.}~\bibnamefont {Cohen}},\ and\ \bibinfo {author} {\bibfnamefont
  {M.}~\bibnamefont {Eckstein}},\ }\bibfield  {title} {\bibinfo {title}
  {Numerically exact simulation of photodoped mott insulators},\ }\href
  {https://doi.org/10.1103/PhysRevLett.132.176501} {\bibfield  {journal}
  {\bibinfo  {journal} {Phys. Rev. Lett.}\ }\textbf {\bibinfo {volume} {132}},\
  \bibinfo {pages} {176501} (\bibinfo {year} {2024})}\BibitemShut {NoStop}%
\bibitem [{\citenamefont {Arrigoni}\ \emph {et~al.}(2013)\citenamefont
  {Arrigoni}, \citenamefont {Knap},\ and\ \citenamefont {von~der
  Linden}}]{ar.kn.13}%
  \BibitemOpen
  \bibfield  {author} {\bibinfo {author} {\bibfnamefont {E.}~\bibnamefont
  {Arrigoni}}, \bibinfo {author} {\bibfnamefont {M.}~\bibnamefont {Knap}},\
  and\ \bibinfo {author} {\bibfnamefont {W.}~\bibnamefont {von~der Linden}},\
  }\bibfield  {title} {\bibinfo {title} {Nonequilibrium dynamical mean field
  theory: an auxiliary quantum master equation approach},\ }\href
  {https://doi.org/10.1103/PhysRevLett.110.086403} {\bibfield  {journal}
  {\bibinfo  {journal} {Phys. Rev. Lett.}\ }\textbf {\bibinfo {volume} {110}},\
  \bibinfo {pages} {086403} (\bibinfo {year} {2013})}\BibitemShut {NoStop}%
\bibitem [{\citenamefont {Meyn}\ and\ \citenamefont
  {Tweedie}(1993)}]{me.tw.93}%
  \BibitemOpen
  \bibfield  {author} {\bibinfo {author} {\bibfnamefont {S.}~\bibnamefont
  {Meyn}}\ and\ \bibinfo {author} {\bibfnamefont {R.}~\bibnamefont {Tweedie}},\
  }\href {https://doi.org/10.1007/978-1-4471-3267-7} {\emph {\bibinfo {title}
  {Markov Chains and Stochastic Stability}}},\ Communications and Control
  Engineering Series\ (\bibinfo  {publisher} {Springer-Verlag},\ \bibinfo
  {address} {London},\ \bibinfo {year} {1993})\BibitemShut {NoStop}%
\bibitem [{\citenamefont {Parcollet}\ \emph {et~al.}(2015)\citenamefont
  {Parcollet}, \citenamefont {Ferrero}, \citenamefont {Ayral}, \citenamefont
  {Hafermann}, \citenamefont {Krivenko}, \citenamefont {Messio},\ and\
  \citenamefont {Seth}}]{pa.fe.15}%
  \BibitemOpen
  \bibfield  {author} {\bibinfo {author} {\bibfnamefont {O.}~\bibnamefont
  {Parcollet}}, \bibinfo {author} {\bibfnamefont {M.}~\bibnamefont {Ferrero}},
  \bibinfo {author} {\bibfnamefont {T.}~\bibnamefont {Ayral}}, \bibinfo
  {author} {\bibfnamefont {H.}~\bibnamefont {Hafermann}}, \bibinfo {author}
  {\bibfnamefont {I.}~\bibnamefont {Krivenko}}, \bibinfo {author}
  {\bibfnamefont {L.}~\bibnamefont {Messio}},\ and\ \bibinfo {author}
  {\bibfnamefont {P.}~\bibnamefont {Seth}},\ }\bibfield  {title} {\bibinfo
  {title} {Triqs: A toolbox for research on interacting quantum systems},\
  }\href {https://doi.org/https://doi.org/10.1016/j.cpc.2015.04.023} {\bibfield
   {journal} {\bibinfo  {journal} {Computer Physics Communications}\ }\textbf
  {\bibinfo {volume} {196}},\ \bibinfo {pages} {398} (\bibinfo {year}
  {2015})}\BibitemShut {NoStop}%
\bibitem [{\citenamefont {Seth}\ \emph {et~al.}(2016)\citenamefont {Seth},
  \citenamefont {Krivenko}, \citenamefont {Ferrero},\ and\ \citenamefont
  {Parcollet}}]{triqscthyb}%
  \BibitemOpen
  \bibfield  {author} {\bibinfo {author} {\bibfnamefont {P.}~\bibnamefont
  {Seth}}, \bibinfo {author} {\bibfnamefont {I.}~\bibnamefont {Krivenko}},
  \bibinfo {author} {\bibfnamefont {M.}~\bibnamefont {Ferrero}},\ and\ \bibinfo
  {author} {\bibfnamefont {O.}~\bibnamefont {Parcollet}},\ }\bibfield  {title}
  {\bibinfo {title} {Triqs/cthyb: A continuous-time quantum monte carlo
  hybridisation expansion solver for quantum impurity problems},\ }\href
  {https://doi.org/https://doi.org/10.1016/j.cpc.2015.10.023} {\bibfield
  {journal} {\bibinfo  {journal} {Computer Physics Communications}\ }\textbf
  {\bibinfo {volume} {200}},\ \bibinfo {pages} {274} (\bibinfo {year}
  {2016})}\BibitemShut {NoStop}%
\bibitem [{\citenamefont {Kraberger}\ \emph {et~al.}(2017)\citenamefont
  {Kraberger}, \citenamefont {Triebl}, \citenamefont {Zingl},\ and\
  \citenamefont {Aichhorn}}]{PhysRevB.96.155128}%
  \BibitemOpen
  \bibfield  {author} {\bibinfo {author} {\bibfnamefont {G.~J.}\ \bibnamefont
  {Kraberger}}, \bibinfo {author} {\bibfnamefont {R.}~\bibnamefont {Triebl}},
  \bibinfo {author} {\bibfnamefont {M.}~\bibnamefont {Zingl}},\ and\ \bibinfo
  {author} {\bibfnamefont {M.}~\bibnamefont {Aichhorn}},\ }\bibfield  {title}
  {\bibinfo {title} {Maximum entropy formalism for the analytic continuation of
  matrix-valued green's functions},\ }\href
  {https://doi.org/10.1103/PhysRevB.96.155128} {\bibfield  {journal} {\bibinfo
  {journal} {Phys. Rev. B}\ }\textbf {\bibinfo {volume} {96}},\ \bibinfo
  {pages} {155128} (\bibinfo {year} {2017})}\BibitemShut {NoStop}%
\bibitem [{\citenamefont {Sachdev}(1999)}]{sachdev}%
  \BibitemOpen
  \bibfield  {author} {\bibinfo {author} {\bibfnamefont {S.}~\bibnamefont
  {Sachdev}},\ }\href@noop {} {\emph {\bibinfo {title} {Quantum Phase
  Transitions}}}\ (\bibinfo  {publisher} {Cambridge University Press},\
  \bibinfo {address} {Cambridge},\ \bibinfo {year} {1999})\BibitemShut
  {NoStop}%
\bibitem [{\citenamefont {Altland}\ and\ \citenamefont
  {Simons}(2010)}]{al.si.10}%
  \BibitemOpen
  \bibfield  {author} {\bibinfo {author} {\bibfnamefont {A.}~\bibnamefont
  {Altland}}\ and\ \bibinfo {author} {\bibfnamefont {B.~D.}\ \bibnamefont
  {Simons}},\ }\href@noop {} {\emph {\bibinfo {title} {Condensed matter field
  theory}}},\ \bibinfo {edition} {2nd}\ ed.\ (\bibinfo  {publisher} {Cambridge
  University Press},\ \bibinfo {year} {2010})\BibitemShut {NoStop}%
\bibitem [{\citenamefont {Balents}(2010)}]{balents.10}%
  \BibitemOpen
  \bibfield  {author} {\bibinfo {author} {\bibfnamefont {L.}~\bibnamefont
  {Balents}},\ }\bibfield  {title} {\bibinfo {title} {Spin liquids in
  frustrated magnets},\ }\href {https://doi.org/10.1038/nature08917} {\bibfield
   {journal} {\bibinfo  {journal} {Nature}\ }\textbf {\bibinfo {volume}
  {464}},\ \bibinfo {pages} {199} (\bibinfo {year} {2010})}\BibitemShut
  {NoStop}%
\bibitem [{\citenamefont {Giamarchi}(2003)}]{giamarchi.03}%
  \BibitemOpen
  \bibfield  {author} {\bibinfo {author} {\bibfnamefont {T.}~\bibnamefont
  {Giamarchi}},\ }\href
  {https://doi.org/10.1093/acprof:oso/9780198525004.001.0001} {\emph {\bibinfo
  {title} {Quantum Physics in One Dimension}}}\ (\bibinfo  {publisher} {Oxford
  University Press},\ \bibinfo {year} {2003})\BibitemShut {NoStop}%
\bibitem [{\citenamefont {Lee}\ \emph {et~al.}(2006)\citenamefont {Lee},
  \citenamefont {Nagaosa},\ and\ \citenamefont {Wen}}]{le.na.06}%
  \BibitemOpen
  \bibfield  {author} {\bibinfo {author} {\bibfnamefont {P.~A.}\ \bibnamefont
  {Lee}}, \bibinfo {author} {\bibfnamefont {N.}~\bibnamefont {Nagaosa}},\ and\
  \bibinfo {author} {\bibfnamefont {X.-G.}\ \bibnamefont {Wen}},\ }\bibfield
  {title} {\bibinfo {title} {Doping a mott insulator: Physics of
  high-temperature superconductivity},\ }\href
  {https://doi.org/10.1103/RevModPhys.78.17} {\bibfield  {journal} {\bibinfo
  {journal} {Rev. Mod. Phys.}\ }\textbf {\bibinfo {volume} {78}},\ \bibinfo
  {pages} {17} (\bibinfo {year} {2006})}\BibitemShut {NoStop}%
\bibitem [{\citenamefont {Blaha}\ \emph {et~al.}(2020)\citenamefont {Blaha},
  \citenamefont {Schwarz}, \citenamefont {Tran}, \citenamefont {Laskowski},
  \citenamefont {Madsen},\ and\ \citenamefont {Marks}}]{Blaha_2020_PAPER}%
  \BibitemOpen
  \bibfield  {author} {\bibinfo {author} {\bibfnamefont {P.}~\bibnamefont
  {Blaha}}, \bibinfo {author} {\bibfnamefont {K.}~\bibnamefont {Schwarz}},
  \bibinfo {author} {\bibfnamefont {F.}~\bibnamefont {Tran}}, \bibinfo {author}
  {\bibfnamefont {R.}~\bibnamefont {Laskowski}}, \bibinfo {author}
  {\bibfnamefont {G.~K.~H.}\ \bibnamefont {Madsen}},\ and\ \bibinfo {author}
  {\bibfnamefont {L.~D.}\ \bibnamefont {Marks}},\ }\bibfield  {title} {\bibinfo
  {title} {{WIEN2k}: {An APW}+lo program for calculating the properties of
  solids},\ }\href {https://doi.org/10.1063/1.5143061} {\bibfield  {journal}
  {\bibinfo  {journal} {The Journal of Chemical Physics}\ }\textbf {\bibinfo
  {volume} {152}},\ \bibinfo {pages} {074101} (\bibinfo {year}
  {2020})}\BibitemShut {NoStop}%
\bibitem [{\citenamefont {Aichhorn}\ \emph {et~al.}(2016)\citenamefont
  {Aichhorn}, \citenamefont {Pourovskii}, \citenamefont {Seth}, \citenamefont
  {Vildosola}, \citenamefont {Zingl}, \citenamefont {Peil}, \citenamefont
  {Deng}, \citenamefont {Mravlje}, \citenamefont {Kraberger}, \citenamefont
  {Martins}, \citenamefont {Ferrero},\ and\ \citenamefont
  {Parcollet}}]{Aichhorn2016}%
  \BibitemOpen
  \bibfield  {author} {\bibinfo {author} {\bibfnamefont {M.}~\bibnamefont
  {Aichhorn}}, \bibinfo {author} {\bibfnamefont {L.}~\bibnamefont
  {Pourovskii}}, \bibinfo {author} {\bibfnamefont {P.}~\bibnamefont {Seth}},
  \bibinfo {author} {\bibfnamefont {V.}~\bibnamefont {Vildosola}}, \bibinfo
  {author} {\bibfnamefont {M.}~\bibnamefont {Zingl}}, \bibinfo {author}
  {\bibfnamefont {O.~E.}\ \bibnamefont {Peil}}, \bibinfo {author}
  {\bibfnamefont {X.}~\bibnamefont {Deng}}, \bibinfo {author} {\bibfnamefont
  {J.}~\bibnamefont {Mravlje}}, \bibinfo {author} {\bibfnamefont {G.~J.}\
  \bibnamefont {Kraberger}}, \bibinfo {author} {\bibfnamefont {C.}~\bibnamefont
  {Martins}}, \bibinfo {author} {\bibfnamefont {M.}~\bibnamefont {Ferrero}},\
  and\ \bibinfo {author} {\bibfnamefont {O.}~\bibnamefont {Parcollet}},\
  }\bibfield  {title} {\bibinfo {title} {Triqs/dfttools: A triqs application
  for ab initio calculations of correlated materials},\ }\href
  {https://doi.org/https://doi.org/10.1016/j.cpc.2016.03.014} {\bibfield
  {journal} {\bibinfo  {journal} {Computer Physics Communications}\ }\textbf
  {\bibinfo {volume} {204}},\ \bibinfo {pages} {200} (\bibinfo {year}
  {2016})}\BibitemShut {NoStop}%
\bibitem [{\citenamefont {Pizzi}\ \emph {et~al.}(2020)\citenamefont {Pizzi},
  \citenamefont {Vitale}, \citenamefont {Arita}, \citenamefont {Blügel},
  \citenamefont {Freimuth}, \citenamefont {G{\'{e}}ranton}, \citenamefont
  {Gibertini}, \citenamefont {Gresch}, \citenamefont {Johnson}, \citenamefont
  {Koretsune}, \citenamefont {Iba{\~{n}}ez-Azpiroz}, \citenamefont {Lee},
  \citenamefont {Lihm}, \citenamefont {Marchand}, \citenamefont {Marrazzo},
  \citenamefont {Mokrousov}, \citenamefont {Mustafa}, \citenamefont {Nohara},
  \citenamefont {Nomura}, \citenamefont {Paulatto}, \citenamefont
  {Ponc{\'{e}}}, \citenamefont {Ponweiser}, \citenamefont {Qiao}, \citenamefont
  {Thöle}, \citenamefont {Tsirkin}, \citenamefont {Wierzbowska}, \citenamefont
  {Marzari}, \citenamefont {Vanderbilt}, \citenamefont {Souza}, \citenamefont
  {Mostofi},\ and\ \citenamefont {Yates}}]{Pizzi2020}%
  \BibitemOpen
  \bibfield  {author} {\bibinfo {author} {\bibfnamefont {G.}~\bibnamefont
  {Pizzi}}, \bibinfo {author} {\bibfnamefont {V.}~\bibnamefont {Vitale}},
  \bibinfo {author} {\bibfnamefont {R.}~\bibnamefont {Arita}}, \bibinfo
  {author} {\bibfnamefont {S.}~\bibnamefont {Blügel}}, \bibinfo {author}
  {\bibfnamefont {F.}~\bibnamefont {Freimuth}}, \bibinfo {author}
  {\bibfnamefont {G.}~\bibnamefont {G{\'{e}}ranton}}, \bibinfo {author}
  {\bibfnamefont {M.}~\bibnamefont {Gibertini}}, \bibinfo {author}
  {\bibfnamefont {D.}~\bibnamefont {Gresch}}, \bibinfo {author} {\bibfnamefont
  {C.}~\bibnamefont {Johnson}}, \bibinfo {author} {\bibfnamefont
  {T.}~\bibnamefont {Koretsune}}, \bibinfo {author} {\bibfnamefont
  {J.}~\bibnamefont {Iba{\~{n}}ez-Azpiroz}}, \bibinfo {author} {\bibfnamefont
  {H.}~\bibnamefont {Lee}}, \bibinfo {author} {\bibfnamefont {J.-M.}\
  \bibnamefont {Lihm}}, \bibinfo {author} {\bibfnamefont {D.}~\bibnamefont
  {Marchand}}, \bibinfo {author} {\bibfnamefont {A.}~\bibnamefont {Marrazzo}},
  \bibinfo {author} {\bibfnamefont {Y.}~\bibnamefont {Mokrousov}}, \bibinfo
  {author} {\bibfnamefont {J.~I.}\ \bibnamefont {Mustafa}}, \bibinfo {author}
  {\bibfnamefont {Y.}~\bibnamefont {Nohara}}, \bibinfo {author} {\bibfnamefont
  {Y.}~\bibnamefont {Nomura}}, \bibinfo {author} {\bibfnamefont
  {L.}~\bibnamefont {Paulatto}}, \bibinfo {author} {\bibfnamefont
  {S.}~\bibnamefont {Ponc{\'{e}}}}, \bibinfo {author} {\bibfnamefont
  {T.}~\bibnamefont {Ponweiser}}, \bibinfo {author} {\bibfnamefont
  {J.}~\bibnamefont {Qiao}}, \bibinfo {author} {\bibfnamefont {F.}~\bibnamefont
  {Thöle}}, \bibinfo {author} {\bibfnamefont {S.~S.}\ \bibnamefont {Tsirkin}},
  \bibinfo {author} {\bibfnamefont {M.}~\bibnamefont {Wierzbowska}}, \bibinfo
  {author} {\bibfnamefont {N.}~\bibnamefont {Marzari}}, \bibinfo {author}
  {\bibfnamefont {D.}~\bibnamefont {Vanderbilt}}, \bibinfo {author}
  {\bibfnamefont {I.}~\bibnamefont {Souza}}, \bibinfo {author} {\bibfnamefont
  {A.~A.}\ \bibnamefont {Mostofi}},\ and\ \bibinfo {author} {\bibfnamefont
  {J.~R.}\ \bibnamefont {Yates}},\ }\bibfield  {title} {\bibinfo {title}
  {Wannier90 as a community code: new features and applications},\ }\href
  {https://doi.org/10.1088/1361-648x/ab51ff} {\bibfield  {journal} {\bibinfo
  {journal} {Journal of Physics: Condensed Matter}\ }\textbf {\bibinfo {volume}
  {32}},\ \bibinfo {pages} {165902} (\bibinfo {year} {2020})}\BibitemShut
  {NoStop}%
\bibitem [{\citenamefont {Mazzocchi}(2025)}]{data}%
  \BibitemOpen
  \bibfield  {author} {\bibinfo {author} {\bibfnamefont {T.~M.}\ \bibnamefont
  {Mazzocchi}},\ }\href {https://doi.org/10.3217/1q2tq-cjv28} {\bibinfo {title}
  {Mixed-configuration approximation for multi-orbital systems out of
  equilibrium}} (\bibinfo {year} {2025})\BibitemShut {NoStop}%
\bibitem [{\citenamefont {Kubo}(1962)}]{kubo.62}%
  \BibitemOpen
  \bibfield  {author} {\bibinfo {author} {\bibfnamefont {R.}~\bibnamefont
  {Kubo}},\ }\bibfield  {title} {\bibinfo {title} {Generalized cumulant
  expansion method},\ }\href@noop {} {\bibfield  {journal} {\bibinfo  {journal}
  {J. Phys. Soc. Jpn.}\ }\textbf {\bibinfo {volume} {17}},\ \bibinfo {pages}
  {1100} (\bibinfo {year} {1962})}\BibitemShut {NoStop}%
\bibitem [{\citenamefont {Shugard}\ and\ \citenamefont
  {Weeks}(1980)}]{sh.we.80}%
  \BibitemOpen
  \bibfield  {author} {\bibinfo {author} {\bibfnamefont {W.~J.}\ \bibnamefont
  {Shugard}}\ and\ \bibinfo {author} {\bibfnamefont {J.~D.}\ \bibnamefont
  {Weeks}},\ }\bibfield  {title} {\bibinfo {title} {Renormalized finite-cluster
  method for lattice models. i. site renormalization and star cluster
  expansions},\ }\href {https://doi.org/10.1103/PhysRevB.22.5245} {\bibfield
  {journal} {\bibinfo  {journal} {Phys. Rev. B}\ }\textbf {\bibinfo {volume}
  {22}},\ \bibinfo {pages} {5245} (\bibinfo {year} {1980})}\BibitemShut
  {NoStop}%
\bibitem [{\citenamefont {Beach}\ \emph {et~al.}(2000)\citenamefont {Beach},
  \citenamefont {Gooding},\ and\ \citenamefont {Marsiglio}}]{be.go.00}%
  \BibitemOpen
  \bibfield  {author} {\bibinfo {author} {\bibfnamefont {K.~S.~D.}\
  \bibnamefont {Beach}}, \bibinfo {author} {\bibfnamefont {R.~J.}\ \bibnamefont
  {Gooding}},\ and\ \bibinfo {author} {\bibfnamefont {F.}~\bibnamefont
  {Marsiglio}},\ }\bibfield  {title} {\bibinfo {title} {Reliable pad{\'e}
  analytical continuation method based on a high-accuracy symbolic computation
  algorithm},\ }\href@noop {} {\bibfield  {journal} {\bibinfo  {journal} {Phys.
  Rev. B}\ }\textbf {\bibinfo {volume} {61}},\ \bibinfo {pages} {5147}
  (\bibinfo {year} {2000})}\BibitemShut {NoStop}%
\end{thebibliography}%

\end{document}